\documentclass[onecolumn,amsmath,showpacs,11pt]{revtex4}
\usepackage{mathrsfs} 
\usepackage{graphicx}
\usepackage{bm}
\usepackage{float}
\usepackage{fancyhdr}
\usepackage{longtable}
\usepackage{hyperref}
\usepackage{subfigure}
\usepackage{epsfig} 
\usepackage{color,soul}
\usepackage{menukeys}
\usepackage{xcolor}
\usepackage{feynman}
\usepackage{amssymb}
\usepackage{braket}
\usepackage{graphicx}
\usepackage{dcolumn}
\usepackage[mathscr]{eucal}
\usepackage{amsmath}
\usepackage{tabularx}
\usepackage{booktabs}
\usepackage{slashed}
\usepackage{mathtools,slashed}
\usepackage[utf8x]{inputenc}

\makeatletter
\def\pslashed#1{%
\expandafter\ifx\csname psla@\string#1\endcsname\relax
{\mathpalette{\sla@/00}{\phantom{#1}}}%
\else
\csname psla@\string#1\endcsname\fi}
\def\declarepslashed#1#2#3#4#5{%
\expandafter\def\csname psla@\string#5\endcsname{%
#1{\mathpalette{\sla@{#2}{#3}{#4}}{\phantom{#5}}}}}
\makeatother
\declarepslashed{}{/}{.08}{0}{D}

\newcommand{\hs}{\hspace*{0.5cm}}

\newcommand{\be}{\begin{equation}}
	\newcommand{\ee}{\end{equation}}
\newcommand{\bea}{\begin{eqnarray}}
	\newcommand{\eea}{\end{eqnarray}}
\newcommand{\ben}{\begin{enumerate}}
	\newcommand{\een}{\end{enumerate}}
\newcommand{\bde}{\begin{widetext}}
	\newcommand{\ede}{\end{widetext}}

\newcommand{\crn}{\nonumber \\}

\newcommand{\al}{\alpha}
\newcommand{\la}{\lambda}

\newcommand{\fr}{\frac}
\newcommand{\bc}{\begin{center}}
	\newcommand{\ec}{\end{center}}
\newcommand{\Ga}{\Gamma}
\newcommand{\ga}{\gamma}

\newcommand{\ph}{\phi}

\begin{document}
	
\title{ New physics in $\text{b} \rightarrow \text{s}$ transitions in the \text{MF331} model}
\author{N. T. Duy $^{a,b}$}
\author{P. N. Thu $^{a,c}$}
\author{D. T. Huong$^{a}$}
\email{dthuong@iop.vast.vn}
\email{ntduy@iop.vast.vn}

\affiliation{$^a$ Institute of Physics, VAST, 10 Dao Tan, Ba Dinh, Hanoi, Vietnam\\
	 $^b$ Graduate University of Science and Technology,
	Vietnam Academy of Science and Technology,
	18 Hoang Quoc Viet, Cau Giay, Hanoi, Vietnam \\
	$^c$ Faculty of Natural Sciences and Technology
	Tay Bac University, Quyet Tam Ward, Son La City, Son La Province}
\date{\today}

\date{\today}

\begin{abstract}
There are two sources that help to explain the $\text{R}_\text{K}$, $\text{R}_{\text{K}^*}$ anomalies in the MF331 model. The first is non-LFUV couplings of the new neutral gauge boson $\text{Z}^{\prime}$ with leptons, $\text{g}^{\text{Z}^\prime}(e)\neq \text{g}^{\text{Z}^\prime}(\mu,\tau)$, which causes the $\text{R}_\text{K}$, $\text{R}_{\text{K}^*}$ anomalies via $\text{Z}^\prime$-penguin diagrams involving newly charged gauge bosons $\text{X}^{\pm}_{\mu}$, and exotic U-quarks. The box diagram's contribution is the second source, which induced only the first lepton generation. We show that the penguin diagrams can not explain $\text{R}_\text{K}$, $\text{R}_{\text{K}^*}$ anomalies, and that the box diagram is required. The experimental constraints for $\text{R}_\text{K}$ and $\text{R}_{\text{K}^*}$ result in new particle mass degeneracy. The contributions of NP to the branching ratios $\text{Br}(\text{B}\to \mu^+ \mu^-), \text{Br}(\text{b}\to s \gamma)$ predict results that agree with the experimental limits in the allowed region of the NP scale.
\end{abstract}

\maketitle
	\section{Introduction} 
In recent years, the LHCb has provided observations that show a conflict between the standard model (SM) predictions and the experimental results. The results of the angular analysis of the decay $\text{B}^0 \rightarrow \text{K}^{0*} \mu^+ \mu^-$ and measurements of the branching fraction of several $\text{b} \rightarrow \text{s} \text{l}^+ \text{l}^-$ \cite{Aaij_2016, Aaboud_2018,Aubert_2006,Lees_2016, Wei_2009,bellecollaboration2016leptonflavordependent,Aaltonen_2012,Aaij:2012vr,Khachatryan_2016,Sirunyan_2018} are in tension with those of SM. Some of these tensions can be explained by the involvement of hadronic uncertainties arising from the different long-distance effects \cite{Khodjamirian_2010,Khodjamirian_2013,Descotes_Genon_2014,Capdevila_2017,Blake_2018}, while the rest are explained by NP signs \cite{Altmannshofer_2009, Bobeth_2012, Matias_2012, Descotes_Genon_2013, Matias_2014}. Lepton flavor universality violating (LFUV) observables, such as the ratios of branching fractions involving both $\text{b} \rightarrow \text{s} \mu^+ \mu^-$ and $\text{b} \rightarrow \text{s} e^+ e^-$ transitions, are also intriguing to theorists. The LHCb and Belle collaborations measured \cite{Aaij_2014,Aaij_2019,Choudhury_2021,Aaij2021} the ratio $\text{R}_{\text{K}}\equiv \frac{\text{Br}\left(\text{B}^{+}\rightarrow \text{K}^{+} \mu^+ \mu^-\right)}{\text{Br}\left(\text{B}^{+}\rightarrow \text{K}^{+} \text{e}^+ \text{e}^-\right)}$ in the low dilepton invariant mass-squared range $\left(1.0 \le \text{q}^2 \le 6.0 \ \text{GeV}^2  \right)$. The LHCb has reported the latest value of $\text{R}_{\text{K}}$ \cite{Aaij2021}, $\text{R}_\text{K}^{\text{LHCb}}\left(\left[1.1, 6 \right] \text{GeV}^2 \right)= 0.846^{+0.042+0.013}_{-0.039-0.012}$, which showed $3.1 \sigma$ deviation from the SM expectation \cite{Bordone_2016, straub2018flavio} of $\simeq 1$, giving evidence for the violation of lepton universality in these decays. Another ratio was reported by the LHCb \cite{Aaij_2017} and Belle \cite{Wehle_2021}, $\text{R}_{\text{K}^{*}}\equiv \frac{\text{Br}\left(\text{B}\rightarrow \text{K}^{*} \mu^+ \mu^-\right)}{\text{Br}\left(\text{B}\rightarrow \text{K}^{*} \text{e}^+ \text{e}^-\right)}$, which is measured in two dilepton invariant mass squared regions \cite{Aaij_2017},
\begin{eqnarray*}
	\text{R}_{\text{K}^{*}}^{\text{LHCb}} = 
	\begin{cases}
		0.66~^{+~0.11}_{-~0.07}\mathrm{\,(\text{stat})}  \pm 0.03\mathrm{\,(\text{syst})}    & \textrm{for } 0.045 < \text{q}^{2} < 1.1~\mathrm{\,Ge\kern -0.1em V^2\!/}c^4 \, , \\
		0.69~^{+~0.11}_{-~0.07}\mathrm{\,(\text{stat})} \pm 0.05\mathrm{\,(\text{syst})}    & \textrm{for } 1.1\phantom{00} < \text{q}^{2} < 6.0~\mathrm{\,Ge\kern -0.1em V^2\!/}c^4 \, .
	\end{cases}
\end{eqnarray*} 
These ratios have been determined to be $2.1$, $2.5$ standard deviations below their SM expectations,respectively  \cite{Bordone_2016, straub2018flavio,Altmannshofer_2017}. Because the hadronic uncertainties are canceled, the LFUV observables $\text{R}_\text{K}$ and $\text{R}_{\text{K}^*}$ are theoretically clean, contrary to observations of the angular and branching fraction of the $\text{b}\rightarrow \text{s} \text{ll}$ decays. As a result, we can certainly infer the presence of NP. These novel metrics have sparked a lot of interest, leading to a slew of model-independent global analyses \cite{Hiller_2004, Hiller_2015, Altmannshofer_2017a, Arbey_2019, Alok_2019, altmannshofer2021new, hurth2021indications}. The majority of these studies revealed  that the LFUV observables $\text{R}_\text{K}$ and $\text{R}_{\text{K}^*}$ may be explained by using the combination of new contributions of  Wilson coefficients (WCs) associated with $\text{V}$ and $\text{A}$ operators. The NP interpretations of the $\text{R}_\text{K},\text{R}_{\text{K}^*}$ anomalies postulate the existence of a new state with tree-level couplings to muons and quarks, namely $\text{Z}^\prime$ vector bosons \cite{Allanach_2016,Altmannshofer_2016}, scalar leptoquarks \cite{gripaios2015composite,Fajfer_2016}.

For addressing model building, it's reasonable to consider what models naturally lead to the LFUV. Extending the SM's symmetry reveals one of the natural candidates for violating the lepton flavor universality (LFU).  In different approaches to extending the SM symmetry, the class of model-based upon the gauge symmetry $\text{SU}(3)_\text{C} \times \text{SU(3)}_\text{L} \times \text{U}(1)_\text{X}\ \text{(3-3-1)}$ \cite{Pisano_1992,PhysRevLett.69.2889,PhysRevD.47.4158,PhysRevD.22.738,Montero_1993,Foot_1994}is known as an attractive proposal. Because this model explains not only the existence of only three fermions, strong CP conservation, and electric charge quantization, but also dark matter, neutrino masses, cosmic inflation, and matter-antimatter asymmetry, all of which are current SM issues. In order to cancel the $\left[ \text{SU}(3)_\text{L} \right]^3$ anomaly, the number of fermion triplets must equal that of the anti-triplet. Traditionally, the arrangement of the particles is one of the quark families that transforms differently from the remaining quark families, while all lepton families transform identically. According to this arrangement, the models predict the tree-level quark FCNCs coupled to $\text{Z}^\prime$, whereas $\text{Z}^\prime$-boson interacts with a pair of the same flavors and strengths as the three lepton families. It means that this approach predicts the lepton flavor universality (LFU) \cite{Buras_2013,Buras_2014A,Gauld_2014,Buras_2016}. In contrast with this setup , the quark and lepton arrangements flip over, creating new versions that are called the flipped $\text{3-3-1}$ (F331) models\cite{Fonseca_2016, Van_Loi_2020}. The FCNCs are coupled to the $\text{Z}^\prime$ swap from quarks into leptons. Therefore, the F331 models break the LFU at the tree level \cite{Huong_2019}, but quark FCNCs induce it at the one-loop level. It naturally provides solutions for explaining the LFUV measuremens in rare $\text{B}$ meson decays. 

 Based on the minimal flipped \text{3-3-1} (MF331) model \cite{Van_Loi_2020},  a version of the F331 models in which scalar multiplets are reduced to a minimum, we explore the $\text{R}_\text{K}$, $\text{R}_{\text{K}^*}$ anomalies from LFUV including the tree-level and the radiative structure of quark flavor-changing interactions. We are looking for NP parameter space regions that sufficiently represent the experimental data on $\text{R}_\text{K}$ and $\text{R}_{\text{K}^*}$. Furthermore, the $\text{Br}\left( \text{B}_\text{s}\rightarrow \mu^+ \mu^-\right)$ is one of the cleanest observables \cite{Amhis:2019ckw} and there is a minor disagreement with SM prediction \cite{Bobeth:2013uxa}. This tension suggests the same direction as the $\text{R}_{\text{K}^*}$ fit's chosen WCs. As a result, we consider whether the parameter space for fixing $\text{R}_{\text{K}^*}$ and $\text{Br}\left( \text{B}_\text{s}\rightarrow \mu^+ \mu^-\right)$ are compatible. Apart from affecting the above observations, NP can also alter the $\text{Br}\left(\text{b} \rightarrow \text{s} \gamma \right)$. Using the parameter space of the above fits, we estimate the role of NP in the $\text{Br}\left( \text{b} \rightarrow \text{s} \gamma \right)$. 

The paper's structure is organized as follows. In Sec. \ref{MF331}, we give a quick summary of the $\text{MF331}$ model. In Sec. \ref{effH}, we examine all of the NP contributions to the WCs associated decay processes caused by $\text{b}- \text{s}$ transitions provide the effective Hamiltonian for these processes. A detailed description of the $\text{R}_\text{K}$, $\text{R}_{\text{K}^*}$ anomalies included in the global fit is given in Sec. \ref{LFUV}. In Secs. \ref{Bsmumu} and Sec. \ref{bsgamma}, we study the NP contributions to the branching ratios of decays, $\text{B}_\text{s}\rightarrow \mu^+ \mu^-$, $\text{b} \rightarrow \text{s} \gamma$, respectively. Finally, we provide our conclusions in Sec. \ref{concl}. 

\section{A Summary of the $\text{MF331}$ model \label{MF331}}
\subsection{Paticle content and mass spectrum of particles}
The F331 model was first pointed out by Renato M. Fonseca and Martin Hirsch \cite{Fonseca_2016}. The model is based on the extended $\text{SU(3)}_\text{C} \times \text{SU(3)}_\text{L}\times \text{U(1)}_\text{N}$ gauge group, in which the first lepton family is discriminated against, while the remaining lepton families and three quark families are in the same representation by the gauge symmetry, $\text{SU(3)}_\text{L}$. The flipped fermion content is free of all gauge anomalies, as specified by \cite{Fonseca_2016} as  
\bea \psi_{\text{1L}}&=& \left(  \begin{array}{ccc} 
	\xi^{+} & \fr{1}{\sqrt{2}}\xi^{0} & \fr{1}{\sqrt{2}} \nu_1 \\
	\fr{1}{\sqrt{2}} \xi^{0} & \xi^{-} & \fr{1}{\sqrt{2}} \text{e}_1 \\
	\fr{1}{\sqrt{2}} \nu_1 & \fr{1}{\sqrt{2}} \text{e}_1 & \text{E}_1 \\
\end{array}\right)_L\sim \left(1,6,-\fr{1}{3}\right),\\
\psi_{\al L}&=&\left(
\begin{array}{c}
	\nu_{\al }\\
	\text{e}_{\al }\\
	\text{E}_{\al}\\
\end{array}
\right)_L\sim \left(1,3,-\fr{2}{3}\right),\\
\text{e}_{\text{aR}} &\sim& (1,1,-1),\hs \text{E}_{\text{aR}}\sim (1,1,-1),\\
\text{Q}_{\text{aL}}&=&\left(
\begin{array}{c}
	\text{d}_\text{a}\\
	-\text{u}_\text{a}\\
	\text{U}_\text{a}\\
\end{array}
\right)_\text{L}\sim \left(3,3^*,\fr{1}{3}\right),\\ 
\text{u}_{\text{aR}} &\sim& (3,1,2/3),\hs \text{d}_{\text{aR} }\sim (3,1,-1/3),\hs \text{U}_{\text{aR}}\sim (3,1,2/3),\eea where $\text{a}=1,2,3$ and $\alpha=2,3$ are family indices.
The Higgs sector in the F331 model is intricate, with three triplets and one sextet that could lead to the dangerous LFV in the Higgs decay. As a result, the MF331 model \cite{Van_Loi_2020} was presented, in which the fermion content is the same as the F331 model but the Higgs component is decreased to two scalar triplets,
\bea &&\rho=\left(
\begin{array}{c}
	\rho^+_1\\
	\rho^0_2\\
	\rho^0_3\\
\end{array}
\right)\sim (1,3,1/3),\hs \chi=\left(
\begin{array}{c}
	\chi^+_1\\
	\chi^0_2\\
	\chi^0_3\\
\end{array}
\right)\sim (1,3,1/3),\eea 
where their vacuum expectation values (VEVs) have a form 
\bea &&\langle\rho\rangle=\fr{1}{\sqrt2}\left(
\begin{array}{c}
	0\\
	v\\
	w'\\
\end{array}
\right),\hs \langle\chi\rangle=\fr{1}{\sqrt2}\left(
\begin{array}{c}
	0\\
	v'\\
	w\\
\end{array}
\right).\label{29dd}\eea
To keep consistency with the SM and small neutrino masses, the VEVs have to be satisfied $u^\prime, w^\prime \ll v \ll w$.
The scalar potential has a simple form \cite{Van_Loi_2020}, 
\bea V&=& \mu_1^2\rho^{\dagger}\rho+\mu_2^2 \chi^{\dagger}\chi+\la_1(\rho^{\dagger}\rho)^2+\la_2(\chi^{\dagger}\chi)^2+\la_3(\rho^{\dagger}\rho)(\chi^{\dagger}\chi)+\la_4(\rho^{\dagger}\chi)(\chi^{\dagger}\rho)\crn
&&+\left[\bar{\mu}^2_3\chi^\dagger \rho +\bar{\la}_5(\chi^\dagger \rho)^2+(\bar{\la}_6\rho^\dagger \rho+\bar{\la}_7\chi^\dagger \chi)\chi^\dagger \rho+H.c.\right].\label{scd}\eea 
The parameters $\bar{\la}$ and $\bar{\mu}_3$ violate $B-L$ while $\la, \mu_{1,2}$ are the $B-L$ conservation, thus $\bar{\la}\ll \la$ and $\bar{\mu}_3\ll \mu_{1,2}$. After spontaneous symmetry breaking, the MF331 model contains the SM-like Higgs boson and two new Higgs fields $\text{H}_1, \text{H}^\prime$. In the limit, $u^\prime, w^\prime \ll v \ll w$, the physical states have a mass as follows
\bea
m^2_\text{H} \simeq \frac{\left(4 \la_1\la_2-\la_3^2 \right)v^2}{2\la_2}, \hs m^2_{\text{H}_1} \simeq 2 \la_2 w^2, \hs m^2_{\text{H}^\prime} \simeq \frac{\la_4}{2}\left(v^2+w^2 \right),
\eea 
and the Higgs triplets , $\rho, \chi$, are presented through the physical states as follows
\be \rho\simeq \left(
\begin{array}{c}
	\text{G}^+_\text{W}\\
	\fr{1}{\sqrt{2}}(v+\text{H}+i \text{G}_\text{Z})\\
	\fr{1}{\sqrt{2}}w^\prime+\text{H}'\\
\end{array}
\right),\hs \chi\simeq \left(
\begin{array}{c}
	\text{G}^+_\text{X}\\
	\fr{1}{\sqrt{2}}v^\prime+\text{G}^0_\text{Y}\\
	\fr{1}{\sqrt{2}}(w+\text{H}_1+i\text{G}_{\text{Z}'})\\
\end{array}
\right),\label{615d}
\ee 
where $\text{G}_{\text{W,X,Y,Z},\text{Z}^\prime}$ are the Goldstone bosons.   
Because the number of Higgs multiplets is reduced, light fermions gain mass through non-standard interactions characterized by dimension-six operators, whereas the masses of heavy quarks and leptons are determined by normal four-dimensional operators. You may find the total Yukawa interactions in up to six dimensions, as well as the fermion mass spectrum in \cite{Van_Loi_2020}. 

To wrap up this part, let's review the key points concerning the gauge bosons sector. Apart from the SM gauge bosons $\text{Z}, \text{W}^\pm$, the MF331 model includes non-Hermitian gauge bosons $\text{X}^\pm, \text{Y}^{0,0*}$, as well as one new neutral gauge boson $\text{Z}^\prime$, all of which have matching masses
\bea
&& m^2_\text{W}\simeq \fr{g^2v^2}{4},\hs m^2_\text{X}\simeq \fr{g^2w^2}{4},\hs m^2_\text{Y}\simeq \fr{g^2(v^2+w^2)}{4}, \nonumber \\ && m^2_\text{Z}\simeq \fr{g^2v^2}{4\text{c}_\text{W}^2},\hs m^2_{\text{Z}^\prime}\simeq \fr{g^2[\text{c}^2_{2\text{W}}v^2+4\text{c}^4_\text{W} w^2]}{4\text{c}^2_\text{W}(3-4\text{s}^2_\text{W})},
\eea
where $\text{c}_\text{W}=\cos \theta_\text{W},\text{s}_\text{W}=\sin \theta_\text{W} $, $\theta_\text{W}$ is the Weinberg angle
which is determined by $\text{s}_\text{W}= \frac{\sqrt{3}t_\text{X}}{\sqrt{3+4t_\text{X}^2}}$ with $\text{t}_\text{X}=\frac{g_\text{X}}{g}$. 
\subsection{Charged and neutral currents}
The interactions of gauge bosons and fermions are derived from the Lagrangian,
\bea
\mathcal{L}_{\text{Fermion}}= \text{i}\bar{\Psi}\ga^\mu \text{D}_\mu \Psi, \label{inter1}
\eea
where $\Psi$ runs on all over the fermion multiplets of the model. The covariant derivative is determined as
$\text{D}_\mu = \partial_\mu+ig_s t_a\text{G}^a_\mu+ig\text{P}_\mu$, where $t_a$ are the generators of $\text{SU}(3)_\text{C}$ group and equal to $0$ for leptons and
$\frac{\la_\text{a}}{2} $ for quarks. $\text{P}_\mu$ contains the generators of $\text{SU}(3)_\text{L} \times \text{U}(1)_\text{N}$ groups. The form of $\text{P}_\mu$ depends on the representations of  $\text{SU}(3)_\text{L}$ group and $\text{U}(1)_{\text{X}}$ charge (X), namely 

\begin{eqnarray*}
	\text{P}_\mu \Psi =  
	\begin{cases}
		\left( \text{T}_\text{a}\text{A}^{\text{a}}_{ \mu}+\text{t}_\text{X} \text{X} \text{B}_\mu \right)\Psi, &  \text{for triplet of } \text{SU}(3)_\text{L}, \\
	\text{A}^{\text{a}}_{ \mu}	\left( \text{T}_\text{a} \Psi+\Psi \text{T}_\text{a} \right)+\text{t}_\text{X} \text{X} \text{B}_\mu \Psi, & \text{for sextet of }\text{SU}(3)_\text{L},
	\end{cases}
\label{dao}\end{eqnarray*} 
where $\text{T}_\text{a}=\frac{\la_a}{2}$, while $\text{T}_\text{a}$ vanish for the right-handed fermion singlets.
In the F331 model, the first lepton family transforms as a sextet of $\text{SU}(3)_\text{L}$, while the remaining two families transform as a triplet, leaving the LFUV in both charged and neutral lepton currents. By substituting $\text{P}_\mu \Psi$ into Eq.(\ref{inter1}), one can obtain the charged current interactions shown as
\bea
\mathcal{L}^{\text{C.C}}= \text{J}_\text{W}^{-\mu}\text{W}^+_\mu+\text{J}_\text{X}^{-\mu}\text{X}^+_\mu+\text{J}_\text{Y}^{0 \mu} \text{Y}^0_\mu+\text{H.c},
\eea   
\bea
\text{J}_{\text{W}}^{-\mu}&&=-\frac{g}{\sqrt{2}}\left\{\bar{\nu}_{\text{aL}}\ga^\mu \text{e}_{\text{aL}}+\bar{\text{u}}_{\text{aL}}\ga^\mu \text{d}_{\text{aL}} +\sqrt{2}\left(\bar{\xi^+_{\text{L}}}\ga^\mu \xi^0_{\text{L}} +\bar{\xi^0_{\text{L}}}\ga^\mu\xi^-_{\text{L}}\right)\right\},\\
\text{J}_{\text{X}}^{-\mu}&&=-\frac{g}{\sqrt{2}}\left\{\bar{\nu}_{\al \text{L}}\ga^\mu \text{E}_{\al \text{L}}+\sqrt{2}\left( \bar{\nu}_{1 \text{L}}\ga^\mu \text{E}_{1 \text{L}}+\bar{\xi^+_{\text{L}}}\ga^\mu \nu_{1\text{L}} \right)+\bar{\xi_{\text{L}}^0}\ga^\mu \text{e}_{1\text{L}}-\bar{\text{U}}_{\text{aL}}\ga^\mu \text{d}_{\text{aL}} \right\}, \label{cc1}\\
\text{J}_{\text{Y}}^{0\mu}&&=-\frac{g}{\sqrt{2}}\left\{\bar{\text{e}}_{\al \text{L}} \ga^\mu \text{E}_{\al \text{L}}+\sqrt{2}\left(  \bar{\text{e}}_{1\text{L}} \ga^\mu \text{E}_{1\text{L}} +\bar{\xi^-_{\text{L}}} \ga^\mu \text{e}_{1\text{L}}\right)+\bar{\xi_{\text{L}}^0} \ga^\mu \nu_{1\text{L}}+\bar{\text{U}}_{\text{aL}}\ga^\mu \text{u}_{\text{aL}}\right\}.
\eea
In the charged currents associated with the new charged gauge bosons, there is a violation of LFU, particularly in the SM charged leptons when the only electron interacts with the $\text{X}^\pm$ boson. This interaction should play a significant role in understanding the $\text{R}_{\text{K}},\text{R}_{\text{K}^*}$ anomalies. 

One can also extract the neutral current pieces from Eq.(\ref{inter1})
\bea
\mathcal{L}^{\text{N.C}} =-\frac{g}{2 \text{c}_\text{W}} \bar{f} \ga^\mu \left\{ g_V^{Z}(f)-g_A^Z(f)\ga_5 \right\}f \text{Z}_\mu
-\frac{g}{2 \text{c}_\text{W}} \bar{f} \ga^\mu \left\{ g_V^{Z^\prime}(f)-g_A^{Z^\prime}(f)\ga_5 \right\}f \text{Z}_\mu^\prime,
\eea
where the couplings $ g_V^{Z,(Z^\prime)}(f), g_A^{Z,(Z^\prime)}(f)$ are taken from {\cite{Van_Loi_2020}. In Tabel (\ref{coupgauge}), we outline typical couplings of neutral gauge bosons with fermions for convenience of future study.The LFUV can be seen clearly in the interactions of  $\text{Z}^\prime$ boson.
	
Closing this section, we would like to point out that three generations of quarks transform uniformly under the $\text{SU(3)}_L \times \text{U(1)}_X$ groups, and FCNCs processes involving $\text{b} \rightarrow \text{s}$ transitions are loop suppressed. In the upcoming sections, these processes will be studied in further depth.
\begin{table}[!h]
	\bc
	\begin{tabular}{|c|c|c|c|c|}
		\hline
		$f$ & $g^{Z'}_V(f)$ & $g^{Z'}_A(f)$ &$g_L^{Z'}(f)$ &$g_R^{Z'}(f)$ \\  
		\hline
		$\text{e}_1$ & $\frac{1-2\text{c}_{\text{2W}}}{2\sqrt{1+2\text{c}_{\text{2W}}}}$ & $-\frac{1}{2\sqrt{1+2\text{c}_{\text{2W}}}}$ & $\frac{-\text{c}_{\text{2W}}}{2\sqrt{1+2\text{c}_{\text{2W}}}} $&$\frac{\text{s}_{\text{W}}^2}{\sqrt{1+2\text{c}_{\text{2W}}}}$\\ 
		\hline
		$\text{e}_\al$ & $\frac{2-\text{c}_{\text{2W}}}{2\sqrt{1+2\text{c}_{\text{2W}}}}$ & $\frac{\text{c}_{\text{2W}}}{2\sqrt{1+2\text{c}_{\text{2W}}}}$& $\frac{1}{2\sqrt{1+2\text{c}_{\text{2W}}}} $&$\frac{\text{s}_{\text{W}}^2}{\sqrt{1+2\text{c}_{\text{2W}}}}$ \\ 
		\hline
		$\text{d}_\text{a}$ &$-\frac{\sqrt{1+2\text{c}_{\text{2W}}}}{6}$ &  $-\frac{1}{2\sqrt{1+2\text{c}_{\text{2W}}}}$ & $-\frac{2+\text{c}_{\text{2W}}}{6\sqrt{1+2\text{c}_{\text{2W}}}} $&$\frac{\text{s}_{\text{W}}^2}{3\sqrt{1+2\text{c}_{\text{2W}}}}$\\
		\hline
		$\text{U}_\text{a}$ &$\frac{7\text{c}_{\text{2W}}-1}{6\sqrt{1+2\text{c}_{\text{2W}}}}$ &  $\frac{\text{c}_{\text{W}}^2}{\sqrt{1+2\text{c}_{\text{2W}}}}$& $\frac{1+5\text{c}_{\text{2W}}}{6\sqrt{1+2\text{c}_{\text{2W}}}} $&$-\frac{2\text{s}_{\text{W}}^2}{3\sqrt{1+2\text{c}_{\text{2W}}}}$\\
		\hline
		$f$ & $g^{Z}_V(f)$ & $g^{Z}_A(f)$&$g_L^{Z}(f)$ &$g_R^{Z}(f)$ \\
		\hline
		$\text{e}_\text{a}$ & $-\fr 1 2 +2\text{s}^2_\text{W}$ & $-\fr 1 2$ & $-\fr{1}{2}+s_{\text{W}}^2$ & $s_{\text{W}}^2$ \\ 
		\hline
		$\text{d}_\text{a}$ &$-\fr 1 2+\fr 2 3 \text{s}^2_\text{W}$ &  $-\fr 1 2$ & $-\fr{1}{2}+\fr{1}{3}s_{\text{W}}^2$ & $\fr{1}{3}s_{\text{W}}^2$ \\
		\hline
		$\text{U}_\text{a}$ &$-\frac{4}{3}\text{s}^2_\text{W}$ &  $0$ & $-\fr{2}{3}s_{\text{W}}^2$ & $-\fr{2}{3}s_{\text{W}}^2$ \\
		\hline 
	\end{tabular}
	\caption{\label{coupgauge} Some couplings of $\text{Z},\text{Z}^\prime$ with fermions.}
	\ec
\end{table}
\section{Effective Hamiltonian for decay processes induced by $\text{b}- \text{s}$ transitions \label{effH} } 
In the \text{MF331} model, the decay processes, $\text{b} \rightarrow \text{s} \gamma$, $\text{B} \rightarrow \text{l} \text{l}$,  $\text{b} \rightarrow \text{s} \text{l} \text{l}$, are governed by the dimension six operators, $\text{O}_{7,8,9,10}$. The relevant effective Hamiltonian can be written in the following form
\bea 
\mathcal{H}_{eff}=-\frac{4 \text{G}_\text{F}}{\sqrt{2}}\text{V}_{\text{ts}}\text{V}^*_{\text{ts}} \times \sum_{\text{i}=7,8,9,10}\left\{\text{C}_\text{i}(\mu)\text{O}_\text{i}(\mu)\right\}+\text{H.c}.,
\label{six1}\eea 
where
\bea
\text{O}_{7} && = \frac{e}{4 \pi^2} m_b\left( \bar{s} \sigma^{\mu \nu}P_R b \right)F_{\mu \nu}, \hs \text{O}_8 = \frac{g_s}{16 \pi^2}m_b \left(\bar{s}_\al \sigma_{\mu \nu}(T_a)^{\al \beta} P_R b_\beta \right) G^{a \mu \nu},\crn
\text{O}_9&& =\frac{e^2}{16 \pi^2}\left(\bar{s}\ga_\mu P_L b\right) \left(\bar{l}\ga^\mu l \right), \hs \text{O}_{10}=\frac{e^2}{16 \pi^2}\left(\bar{s}\ga_\mu P_L b\right) \left(\bar{l}\ga^\mu  \ga_5 l \right).\eea

The MF331 model does not predict the existence of tree-level FCNCs in the quark sector because all quark families are identical transformations under the $\text{SU}(3)_\text{L}$ group, but it does allow them at the loop level. As a result, the one-loop adjustments determine the transition $\text{b} \rightarrow \text{s}$. For convenience, the WCs are divided into the following contributions
\bea
\text{C}_7 =\text{C}_7^{\text{eff-SM}}+\Delta \text{C}_7, \hs  \text{C}_8 =\text{C}_8^{\text{SM}}+\Delta \text{C}_8, \hs   \text{C}_{9}=\text{C}_9^{\text{eff-SM}}+ \Delta \text{C}_9, \hs \text{C}_{10}= \text{C}_{10}^{\text{SM}}+ \Delta \text{C}_{10}. 
\label{WS1}\eea

where $\text{C}_{7,9}^{\text{eff-SM}}$ and $\text{C}_{8,10}^{\text{SM}}$ are determined by interactions of the SM  \cite{Du:2015tda,Beneke:2017vpq}, $\Delta \text{C}_{7,8,9,10}^{\text{NP}}$ are determined by the new interactions. The NP contributions to $\Delta \text{C}_{9,10}$ for the first lepton family are completely different from the two other families
because the first generation of leptons transforms differently than the subsequent lepton generations. The first generation, $\Delta C_9^{e}$, gets contributions from the $\gamma,\text{ Z}, \text{Z'}-$penguin and box diagrams, whereas $ \Delta C_9^{\mu, \tau}$ only get contributions from the $\gamma,\text{Z},\text{Z'}$-penguin diagrams. To be explicit,  NP's contributions to the above-mentioned WCs are split as follows:

\bea
\Delta C_{9,10}^e && = \Delta C_{9,10}^{e, \gamma}+ \Delta C_{9,10}^{e, Z}+\Delta C_{9,10}^{e, Z^\prime}+ \Delta C_{9,10}^{e, \text{box}}, \crn \Delta C_{9,10}^{\mu \left(\tau \right)} && = \Delta C_{9,10}^{\mu \left(\tau \right), \gamma}+ \Delta C_{9,10}^{\mu \left(\tau \right), Z}+\Delta C_{9,10}^{\mu \left(\tau \right), Z^\prime}, \crn
\Delta C_{7,8} && = \Delta C_{7,8}^{X}.\eea 
The contribution of each style of diagram is indicted by the superscripts.  Without QCD correction, all NP contributions are calculated in leading order. The $Z-$penguin diagrams are presented in Fig. (\ref{Zpenguin}), which are induced by quark currents coupled to the new charge gauge bosons $X_\mu^\pm$.
 
\begin{figure}[H]	
	\centering
	\includegraphics[height=7cm]{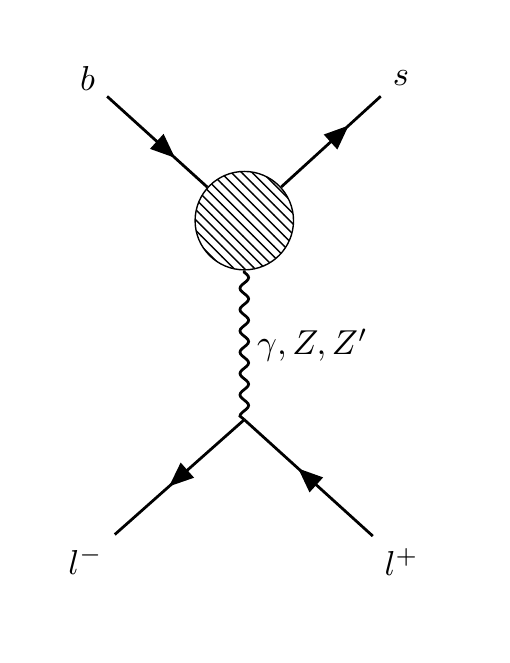} 
	\caption{ $\ga,\text{Z},\text{Z}^\prime$-penguin diagrams induced by new charged gauge boson $\text{X}^{\pm}_{\mu}$. The blob denotes the combination of boson $X^{\pm}$ and new quark $U$ inside the loop. \label{Zpenguin}}	
\end{figure}
Applying the Feynman rules for diagrams given in (\ref{Zpenguin}), we derive the radiative \text{FCNC} coupling $\bar{b}sZ$. Combining the finding with the neutral current of leptons coupled to the $\text{Z}-$boson yields the result
\bea
&& \Delta \text{C}_9^{ Z} = \fr{c_{2W}-2s_W^2}{s_W^2}\left[\fr{-3x^2c_W^2}{8(x-1)^2}\ln{x}+\fr{(x^2+5x-3)c_{2W}+3}{16(x-1)}\right] ,\crn  && \Delta \text{C}_{10}^{Z} =-\fr{1}{s_W^2}\left[\fr{-3x^2c_W^2}{8(x-1)^2}\ln{x}+\fr{(x^2+5x-3)c_{2W}+3}{16(x-1)}\right], 
\eea
where $x=\fr{m_\text{U}^2}{m_\text{X}^2}$. For three generations of leptons, these contributions are the same. It should be noted that the mass of the third exotic quark generation is assumed to be $\text{m}_{\text{U}_3}\gg\text{m}_{\text{U}_{1,2}}$, and similarly in the SM'quarks $\text{m}_\text{t}\gg \text{m}_\text{u,c}$, so the diagrams governed by $\text{U}_3$ dominate.

Because the first generation of leptons transforms differently from the last two generations under the $\text{SU(3)}_\text{L}$  group, the $Z^\prime$ gauge boson interacts with them in a fundamentally different way than the other two generations. These interactions, in combination with the radiative couplings induced by Feynman diagrams (\ref{Zpenguin}), give the contribution  to the WCs. We receive following contributions for different generations:
\bea
&& \Delta \text{C}_9^{e, \text{Z}^\prime} =-\fr{1}{\sqrt{1+2c_{\text{2W}}}s_\text{W}^2}\tilde{g}_\text{V}^{\text{Z}'}(e)\fr{m_Z^2}{m_{\text{Z}'}^2}\left[\fr{x(3x+2)c_\text{W}^2}{2(x-1)^2}\ln{x}+\fr{(3x^2-x-12)c_{\text{2W}}+2(x^2-3x-3)}{8(x-1)}\right]  ,\crn 
&& \Delta \text{C}_{10}^{e, \text{Z}^\prime} =\fr{1}{\sqrt{1+2c_{\text{2W}}}s_\text{W}^2}\tilde{g}_\text{A}^{\text{Z}'}(e)\fr{m_\text{Z}^2}{m_{\text{Z}'}^2}\left[\fr{x(3x+2)	
	c_\text{W}^2}{2(x-1)^2}\ln{x}+\fr{(3x^2-x-12)c_{\text{2W}}+2(x^2-3x-3)}{8(x-1)}\right] , \crn
&& \Delta \text{C}_9^{\mu \left(\tau \right),\text{Z}^\prime} =-\fr{1}{\sqrt{1+2c_{\text{2W}}}s_\text{W}^2}\tilde{g}_\text{V}^{\text{Z}'}(\mu)\fr{m_\text{Z}^2}{m_{\text{Z}'}^2}\left[\fr{x(3x+2)c_\text{W}^2}{2(x-1)^2}\ln{x}+\fr{(3x^2-x-12)c_{\text{2W}}+2(x^2-3x-3)}{8(x-1)}\right]  , \crn
&& \Delta \text{C}_{10}^{\mu \left(\tau \right),\text{Z}^\prime} =\fr{1}{\sqrt{1+2c_{\text{2W}}}s_W^2}\tilde{g}_\text{A}^{\text{Z}'}(\mu)\fr{m_\text{Z}^2}{m_{\text{Z}'}^2}\left[\fr{x(3x+2)c_\text{W}^2}{2(x-1)^2}\ln{x}+\fr{(3x^2-x-12)c_{\text{2W}}+2(x^2-3x-3)}{8(x-1)}\right],\crn
\eea
where $\tilde{g}_{\text{V,A}}^{\text{Z}'}(\text{f})$ are defined in mass eigenstates as 
\bea
&& \tilde{g}_{\text{V}}^{\text{Z}'}(\text{f})= \text{V}_{\text{lL}}^{\dagger}g_\text{L}^{\text{Z}'}(\text{f}) \text{V}_{\text{lL}}+\text{V}_{\text{lR}}^{\dagger}g_\text{R}^{\text{Z}'}(\text{f}) \text{V}_{\text{lR}}, \crn
&& \tilde{g}_{\text{A}}^{\text{Z}'}(\text{f})= \text{V}_{\text{lL}}^{\dagger}g_\text{L}^{\text{Z}'}(\text{f}) \text{V}_{\text{lL}}-\text{V}_{\text{lL}}^{\dagger}g_\text{R}^{\text{Z}'}(\text{f}) \text{V}_{\text{lR}}
\eea
with $g_{\text{L,R}}^{\text{Z}'}$(\text{f}) are the flavor basis couplings of new $Z'$ boson with a pair of left (right) leptons as defined in Table \ref{coupgauge}. The left-handed lepton mixing matrix is $\text{V}_{\text{lL}}$ and it is assumed that $\text{V}_\text{L}=\text{V}_{\text{PMNS}}$, where $\text{V}_{\text{PMNS}}$ denotes \text{Pontecorvo}$-$\text{Maki}$-$\text{Nakagawa}$-$\text{Sakata} (PMNS) matrix. As shown in Table \ref{coupgauge}, the right coupling of $Z'$ with the first lepton generation must be equal to the right coupling of $Z'$ with the two remaining lepton generations, $\text{g}_\text{R}^{Z'} (\text{e}_1)=\text{g}_{\text{R}}^{Z'}(\text{e}_{\al})$. This is explained by the fact that  all three right- handed lepton generations transform identically under $\text{SU}(3)_{\text{L}}$ group, $\text{e}_{\text{aR}}\sim (1,1,-1)$. As a result, the effects of the new right-handed leptons mixing matrix $\text{V}_{\text{lR}}$ will be eliminated, and we will have 
\bea
&& \tilde{g}_{\text{V}}^{\text{Z}'}(\text{f})= \text{V}_{\text{lL}}^{\dagger}g_\text{L}^{\text{Z}'}(\text{f}) \text{V}_{\text{lL}}+g_\text{R}^{\text{Z}'}(\text{f}), \crn
&& \tilde{g}_{\text{A}}^{\text{Z}'}(\text{f})= \text{V}_{\text{lL}}^{\dagger}g_\text{L}^{\text{Z}'}(\text{f}) \text{V}_{\text{lL}}-g_\text{R}^{\text{Z}'}(\text{f}).
\eea
The different arrangement of the fermion generations also leads to a distinct contribution to the WCs. Only $\text{C}_{9,10}^{e}$ benefits from the box diagrams in Fig.(\ref{box}). These additional contributions are given as
\bea
&& \Delta\text{C}_{9}^{e, \text{box}}=-\fr{1}{s_W^2}\fr{m_W^2}{m_X^2}\left\{\fr{x^2[4+(x-8)y]}{16(y-x)(x-1)^2}\ln{x}-\fr{xy[(y-4)^2-12]}{16(y-x)(y-1)^2}\ln{y}+\fr{x(-4+7y)}{16(y-1)(x-1)}\right\}, \crn
&& \Delta \text{C}_{10}^{e, \text{box}}=\fr{1}{s_W^2}\fr{m_W^2}{m_X^2}\left\{\fr{x^2[4+(x-8)y]}{16(y-x)(x-1)^2}\ln{x}-\fr{xy[(y-4)^2-12]}{16(y-x)(y-1)^2}\ln{y}+\fr{x(-4+7y)}{16(y-1)(x-1)}\right\},
\eea
where $y=\fr{m_{\xi^0}^2}{m_\text{X}^2}$.
\begin{figure}[H]
	\centering	
	\includegraphics[height=3.0cm]{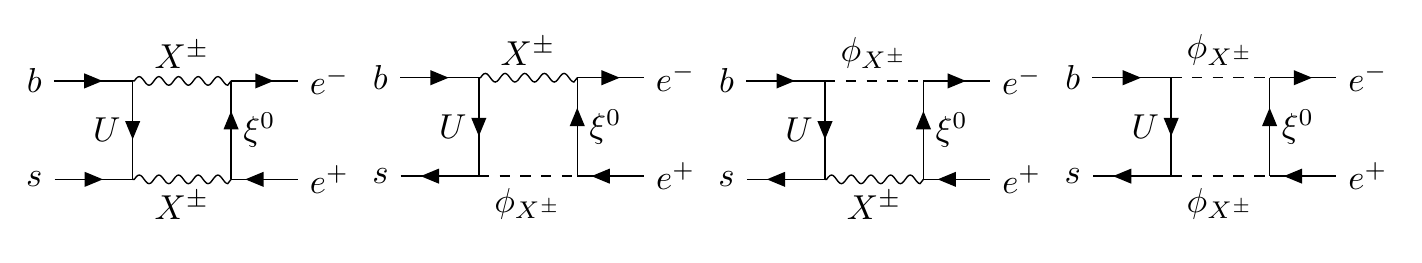} 	
	\caption{\label{box} Box diagrams induced for only the first lepton generation}	
\end{figure} 

The radiative coupling $\text{bs}\gamma$ is produced by the photon penguin diagrams caused by new charged bosons $\text{X}^\pm_\mu$ seen in Fig. (\ref{Zpenguin}). The electromagnetic currents of leptons combine with this coupling to produce additional contributions to $\text{C}_{7,9}$. The outcomes are listed as follows:
\bea
&& \Delta \text{C}_{9}^{ \gamma}=\fr{4}{9}\ln{x}-\fr{x^2(5x^2-2x-6)}{18(x-1)^4}\ln{x}-\fr{-19x^3+25x^2}{36(x-1)^3}, \crn
&& \Delta \text{C}_7^{X}  =-\fr{8x^3+5x^2-7x}{24(x-1)^3}-\fr{x^2(2-3x)}{4(x-1)^4}\ln{x}. 
\eea   
\section{\label{LFUV}Lepton non-universality in $\text{b} \rightarrow \text{s} \text{l}^+ \text{l}^-$}
\subsection{Lepton non-universality in $\text{B}^+ \rightarrow \text{K}^+ \text{l}^+ \text{l}^-$}
As previously stated, the electroweak couplings of charged leptons are distinct in the MF331 model, and as a result, the decay properties of each lepton flavor are expected to be different (referred to as lepton flavor non-universality). 

 In a suitably specified range of the dilepton mass squared, the branching ratios for $\text{B}^{+ \left(* \right)} \rightarrow \text{K}^{+\left( *\right)}\text{l}^+ \text{l}^-$ decays can be expected. The differential branching fraction for  $\text{B}^+ \rightarrow \text{K}^+ \text{l}^+ \text{l}^- $ decays, keeping the lepton mass $\left (\text{m}_\text{l} \right)$, has been studied by \cite{Altmannshofer2015} and given as follows
\bea 
\fr{\text{d}^2\Ga\left(\text{B}^+ \rightarrow \text{K}^+ \text{l}^+ \text{l}^- \right)}{\text{d}\text{q}^2\text{d}(\cos{\theta})}&=&\text{a}(\text{q}^2)
+\text{b}(\text{q}^2)\cos{\theta}+\text{c}(\text{q}^2)\cos^2{\theta}, \label{Br1}
\eea 
where 
\bea
&&\text{a}=\fr{\Ga_0 \la^{3/2}\beta_l}{4} \left\{|\text{G}|^2+|\left(\text{C}_{10}^{\text{SM}}+\Delta C_{10}\right)\text{f}_{+}(\text{q}^2)|^2+|(\text{C}_{10}^{\text{SM}}+\Delta \text{C}_{10})\text{f}_\text{0}(\text{q}^2)|^2\fr{4\text{m}_\text{l}^2}{\la\text{q}^2}\left(\text{m}_\text{B}^2-\text{m}_\text{K}^2\right)^2\right\},\crn 
&&\text{b}=0, \crn
&& \text{c}=-\fr{\Ga_0\la^{3/2}\beta_l^2}{4}\left\{|\text{G}|^2+|\left(\text{C}_{10}^{\text{SM}}+\Delta \text{C}_{10}\right )\text{f}_{+}(\text{q}^2)|^2\right\}\crn \eea
with
\bea
&&\text{G}=\left(\text{C}_9^{\text{eff-SM}}+\Delta \text{C}_9\right)\text{f}_{+}\left(\text{q}^2\right)+\fr{2\text{m}_\text{b}}{\text{m}_\text{B}+\text{m}_\text{K}}\left(\text{C}_7^{\text{eff-SM}}+\Delta \text{C}_7\right)\text{f}_\text{T}(\text{q}^2)+\text{h}_\text{K}(\text{q}^2), \crn 
&&\Ga_0=\fr{\text{G}_\text{F}^2\al_{\text{em}}^2|\text{V}_{\text{tb}}\text{V}_{\text{ts}}^{*}|^2}{512\pi^5\text{m}_\text{B}^3}, \crn
&&\beta_l=\sqrt{1-\fr{4\text{m}_\text{l}^2}{\text{q}^2}},  \eea
$\la$ is a function that depends on the mass of \text{B-}meson, Kaon, and dilepton. It has the following form as
\bea
&&\la \equiv \la(\text{m}_\text{B}^2,\text{m}_\text{K}^2,\text{q}^2)=\text{m}_\text{B}^4+\text{m}_\text{K}^4+\text{q}^4-2
(\text{m}_\text{B}^2\text{m}_\text{K}^2+\text{m}_\text{B}^2\text{q}^2+\text{m}_\text{K}^2\text{q}^2).
\eea
The angle $\theta$ is known as the angle between the negative lepton direction and the $\text{B}$ direction in the $l^+ l^-$ rest frame. 
The effective WCs $\text{C}_{7,9}^{\text{eff-SM}}(\text{q}^2)$ are obtained via the SM's WCs and have the form given  in \cite{Du:2015tda}:  
\bea 
&& \text{C}_7^{\text{eff-SM}}(\text{q}^2)=\text{C}_7-\fr{1}{3}\left(\text{C}_3
+\fr{4}{3}\text{C}_4+20\text{C}_5+\fr{80}{3}\text{C}_6\right)- \fr{\al_s}{4\pi}\left\{\left(\text{C}_1-6\text{C}_2\right)\text{F}_{1,c}^{(7)}(\text{q}^2)+\text{C}_8\text{F}_8^{(7)}(\text{q}^2)\right\} \crn 
&& \text{C}_9^{\text{eff-SM}}(q^2)=\text{C}_9+\fr{4}{3}\left(\text{C}_3+\fr{16}{3}\text{C}_5+\fr{16}{9}\text{C}_6\right)-\text{h}(0,\text{q}^2)\left(\fr{1}{2}\text{C}_3+\fr{2}{3}\text{C}_4+8\text{C}_5+\fr{32}{3}\text{C}_6\right)\crn &&-\text{h}(\text{m}_\text{b}^{\text{pole}},\text{q}^2)\left(\fr{7}{2}\text{C}_3+\fr{2}{3}\text{C}_4+38\text{C}_5+\fr{32}{3}\text{C}_6\right)+\text{h}(\text{m}_\text{c}^{\text{pole}},\text{q}^2)\left(\fr{4}{3}\text{C}_1+\text{C}_2+6\text{C}_3+60\text{C}_5\right) \crn 
&&-\fr{\al_s}{4\pi}\left(\text{C}_1\text{F}_{1,c}^{(9)}(\text{q}^2)+\text{C}_2\text{F}_{2,\text{c}}^{(9)}(\text{q}^2)+\text{C}_8 \text{F}_{8}^{(9)}(\text{q}^2)\right),
\eea 
where the SM's WCs, $\text{C}_{\text{i}}$, are listed in the Table \ref{input1}. The functions $\text{h}(\text{m}_{\text{c,b}}^{\text{pole}},\text{q}^2),\ \text{F}_{1,2,\text{c}}^{(7),(9)}$, and $\text{F}_{8}^{(7),(9)}$ are  determined in \cite{Du:2015tda}.

For $\text{B}\rightarrow \text{K}$ processes, the QCD form factors, $\text{f}_{+,T}(\text{q}^2)$, and $\text{f}_{0}(\text{q}^2)$ can be expressed in the form of a simplified $z$-series expansion as given by \cite{Bourrely:2008za}
\bea 
&& \text{f}^{+,\text{T}}_{\text{B} \rightarrow \text{K}}(\text{q}^2)=\fr{\text{f}^{+,\text{T}}_{\text{B}\rightarrow \text{K}}(0)}{1-\frac{\text{q}^2}{\text{m}_{\text{B}}^2}}\left\{1+\sum_{\text{k}=1}^{\text{N}-1} \text{b}^{+,\text{T}}_{\text{k,K}}\left\{\text{z}(\text{q}^2,\text{t}_0)^\text{k}-\text{z}(0,\text{q}^2)^\text{k}-(-1)^{\text{N-k}}\fr{\text{k}}{\text{N}}\left(\text{z}(\text{q}^2,\text{t}_0)^\text{N}-\text{z}(0,\text{t}_0)^\text{N}\right)\right\}\right\}, \crn 
&& \text{f}^0_{\text{B}\to \text{K}}(\text{q}^2)=\text{f}^{0}_{\text{B}\rightarrow \text{K}}(0)\left\{1+\sum_{\text{k}=1}^{\text{N}} \text{b}^{0}_{\text{k,K}}\left(\text{z}(\text{q}^2,\text{t}_0)^\text{k}-\text{z}(0,\text{t}_0)^\text{k}\right)\right\},
\eea 
where $\text{z}(\text{q}^2,\text{t}_0)=\fr{\sqrt{\text{t}_{+}-\text{q}^2}-\sqrt{\text{t}_{+}-\text{t}_0}}{\sqrt{\text{t}_{+}-\text{q}^2}+\sqrt{\text{t}_{+}-\text{t}_0}}$  with $\text{t}_{+}=(\text{m}_\text{B}+\text{m}_\text{K})^2,\text{t}_0=(\sqrt{\text{m}_\text{B}}-\sqrt{\text{m}_\text{K}})^2(\text{m}_\text{B}+\text{m}_\text{K})$. In this case, $\text{N}=2$ for the vector and tensor form factors $\text{f}_{+,T}(\text{q}^2)$, and $\text{N=1}$ for the scalar form factor $\text{f}_{0}(\text{q}^2)$. The parameters $\text{b}_{\text{k,K}}^{0,\left( +,\text{T}\right)}$ in these form factors have numerical values taken from \cite{Lu:2018cfc}. $\text{h}_\text{K}$ presents the non-factorizable contributions from the weak effective Hamiltonian and has a parameterized form as determined by \cite{Altmannshofer2015}
\bea \text{h}_\text{K}=\text{C}_9^{\text{eff-SM}}(\text{q}^2)\text{f}_{+}(\text{q}^2)\times \left\{ \begin{array}{ll}
	\text{a}_\text{K}\text{e}^{i\phi_\text{a}}+\text{b}_\text{K}\text{e}^{i\phi_\text{b}}(\frac{\text{q}^2}{6} \ \text{GeV}^2) \hs \text{for} \hs \text{q}^2<6 \ \text{GeV}^2 &\\
	\text{c}_{\text{K}}e^{i\phi_c} \hs \hs \hs \hs \hs \hs \hs \hs \text{for} \hs \text{q}^2>6 \ \text{GeV}^2,\
\end{array} \right. \eea
where the strong phases are in the range $\phi_{\text{a,b,c}}\in (-\pi,\pi]$. The coefficients a, b, c satisfy the following conditions: $\text{a} \in [0,0.02]$, $\text{b} \in [0,0.05]$, and $c\in [0,0.05]$.  After subtracting $\theta$ from Eq. (\ref{Br1}), we get 
\bea
\fr{\text{d}\Ga(\text{B}^{+}\rightarrow \text{K}^{+}\text{l}^{+}\text{l}^{-})}{\text{d}\text{q}^2}&&=\fr{\Ga_0\la^{3/2}\beta_\text{l}^3}{3}\left\{\left|\text{G}\right|^2
+\left|\left(\text{C}_{10}^{\text{SM}}+\Delta \text{C}_{10}\right)\text{f}_{+}(\text{q}^2)\right|^2\right\}\crn && +\fr{\Ga_0\la^{3/2}\beta_\text{l}\left(1-\beta_\text{l}^2\right)}{2}\left\{|\text{G}|^2+\fr{(\text{m}_\text{B}^2-\text{m}_\text{K}^2)^2}{\la}\left|\left(\text{C}_{10}^{\text{SM}}+\Delta \text{C}_{10}\right)\text{f}_{0}(\text{q}^2)\right|^2\right\} .\crn \label{Br2}
\eea
The LHCb experiment measures the ratio \cite{Aaij2021} $\text{R}_\text{K}^{\text{LHCb}}\left(\left[1.1, 6 \right] \text{GeV}^2 \right)= 0.846^{+0.042+0.013}_{-0.039-0.012}$, which exhibits $3.1 \sigma$ tension with the SM prediction, as described in the introduction. The LFUV interactions are included in the MF331 model, which may provide a better fit for this data. When it comes to fitting, we use the
$\text{R}_{\text{K}}$ measurement directly, as seen below
\bea
\text{R}_\text{K}=\fr{\int^{\text{q}^2_{\text{max}}}_{\text{q}^2_{\text{min}}}\fr{\text{d}\Ga(\text{B}^{+}\rightarrow \text{K}^{+}\mu^{+}\mu^{-})}{\text{d}\text{q}^2}\text{d}\text{q}^2}{\int^{\text{q}^2_{\text{max}}}_{\text{q}^2_{\text{min}}}
	\fr{\text{d}\Ga(\text{B}^{+}\rightarrow \text{K}^{+}\text{e}^{+}\text{e}^{-})}{\text{dq}^2}\text{dq}^2}.
\eea
 Table \ref{input1} lists the input known parameters as well as the SM's WCs calculated at the next-to-next-to-leading order (NNLO).
\begin{table}
	\begin{center}
		\begin{tabular}{ |c|c|c|c|c| } 
			\hline
			Parameters & Values  &	$\text{C}_\text{i}(\mu=5 \ \text{GeV})$ & Values  \\ 
			\hline 
			$\text{G}_\text{F}$ & 1.166379$\times 10^{-5}\ \text{GeV}^2$ \cite{pdg2020}&	$\text{C}_1$ & -0.25   \\
			$\text{s}_{\text{W}}^{2}$ & 0.23126(5) \text{MeV}  \cite{pdg2020}& $\text{C}_2$ & 1.01 \\
			$\text{m}_\text{Z}$ & $91.1876(21) \text{GeV}$  \cite{pdg2020}& $\text{C}_{3}$ & -0.005 \\
				$\al_{\text{s}}(\text{m}_{\text{Z}})$ & 0.1181(10) \cite{pdg2020} &$\text{C}_4$ & -0.077 \\
			$\al_{\text{em}}(\text{m}_\text{Z})$ & 1/127.955(10) \cite{pdg2020}& $\text{C}_5$ & 0.0003\\
			$|\text{V}_{\text{tb}}\text{V}_{\text{ts}}^{*}|$ & $0.0397^{+0.0008}_{-0.0006}$  \cite{pdg2020}&$\text{C}_6$ & 0.0009   \\
			$\text{m}_{\text{B}^{+}}$ & 5279.34(12) \text{MeV}   \cite{pdg2020}&$\text{C}_7$ & -0.324   \\
			$\text{m}_{\text{K}^{+}}$ & 493.677(16) \text{MeV}   \cite{pdg2020}&$\text{C}_8$ & -0.176     \\ 
			$\text{m}_{\text{b}}^{\text{pole}}$ & 4.91(12) \text{GeV}  \cite{Ali:2013zfa}& $\text{C}_9$ & 4.344   \\
			$\text{m}_{\text{c}}^{\text{pole}}$ & 1.77(14)\text{GeV} \cite{Ali:2013zfa} &$\text{C}_{10}$ & -4.198  \\
			$\text{m}_{\text{B}^0}$ & 5279.65(12) \ \text{MeV}\cite{pdg2020} && \\
			$\text{m}_{\text{K}^{0*}}$ & 895.55(20)\ \text{MeV} \cite{pdg2020} && \\
				$\text{m}_{\text{B}_{\text{s}}}$ & 5366.88(14)\ \text{MeV} \cite{pdg2020} && \\
					$\tau_{\text{B}_{\text{s}}}$ & $1.516(6)\times 10^{-12}$ \text{s}  \cite{pdg2020} && \\
						$\text{f}_{\text{B}_{\text{s}}}$ & 230.3(1.3)\ \text{MeV} \cite{pdg2020} && \\
			$\text{m}_{\text{e}}$ & 0.510998461(31)\ \text{MeV} \cite{pdg2020} && \\
			$\text{m}_{\mu}$ & 105.6583745(24)\ \text{MeV} \cite{pdg2020} && \\
			\hline 
		\end{tabular}
		\caption{\label{input1} Several SM input parameters are being used for numerical study. The SM's WCs, $\text{C}_{1-10}$ at the scale $\mu=5$ GeV, up to NNLO accuracy \cite{Beneke:2017vpq},\cite{MunirBhutta:2020ber}}
	\end{center}
\end{table}
In the \text{MF331} model, the $\text{R}_\text{K}$ also depends on the unknown parameters such as the mass of new particles $\text{m}_{\text{U}_{\text{a}}}, \text{m}_{\zeta^0}$ and $m_{Z'}$. In this paper, we consider two scenarios in which the masses of new fermions $m_U,m_{\zeta^0}$ are either degenerate or non-degenerate
\begin{itemize}

\item Case 1: Degenerate masses	
\bea
 \Delta m\equiv m_{U,Z'}-m_{\zeta^0}=\delta , \hs \delta\ll 1.\eea 
\item Case 2: Non-degenerate masse
\bea m_U=a_1m_{\zeta^0}, \hs m_{Z'}=a_2m_{\zeta^0}, \hs a_{1,2}\sim \mathcal{O}(1). \eea  
\end{itemize}
\begin{figure}[H]
	\centering
		\includegraphics[width=0.7\textwidth]{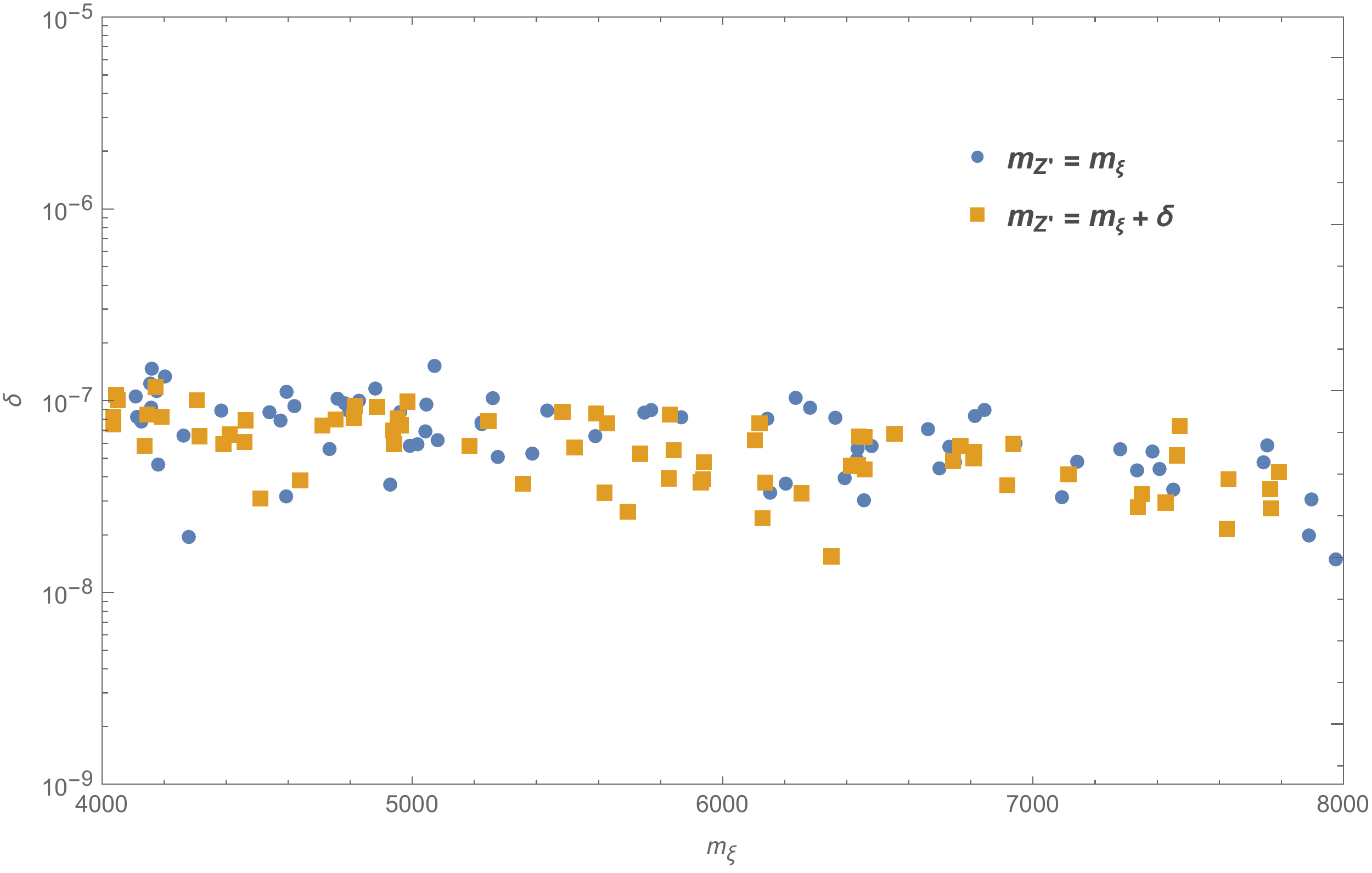}
	\caption{\label{RKa} The frame shows viable parameter space obtained from 
		the most recent measurement \cite{Aaij2021}, $\text{R}_\text{K}^{\text{LHCb}}\left(\left[1.1, 6 \right] \text{GeV}^2 \right)= 0.846^{+0.042+0.013}_{-0.039-0.012}$.} 
\end{figure}

In the first scenario, we seed the values of $\text{m}_{\zeta^0}$ and $\delta $ at random to obtain values that satisfy the $\text{R}_\text{K}$ constraint. The values of $\text{m}_{\zeta^0}$ and $\delta $ are in the range $[4000,8000]$ GeV and $[10^{-8},10^{-5}]$, respectively. Fig. (\ref{RKa}) depicts the acquired results. We see that there is almost no difference between the two kinds of dots, and the allowed range of $\delta$ are from $10^{-6}-10^{-7}$. The finding $\text{R}_\text{K}$ is not overly sensitive to the mass domain of new particles in this region of delta.  
\begin{figure}[H]
	\centering
	\includegraphics[width=0.7\textwidth]{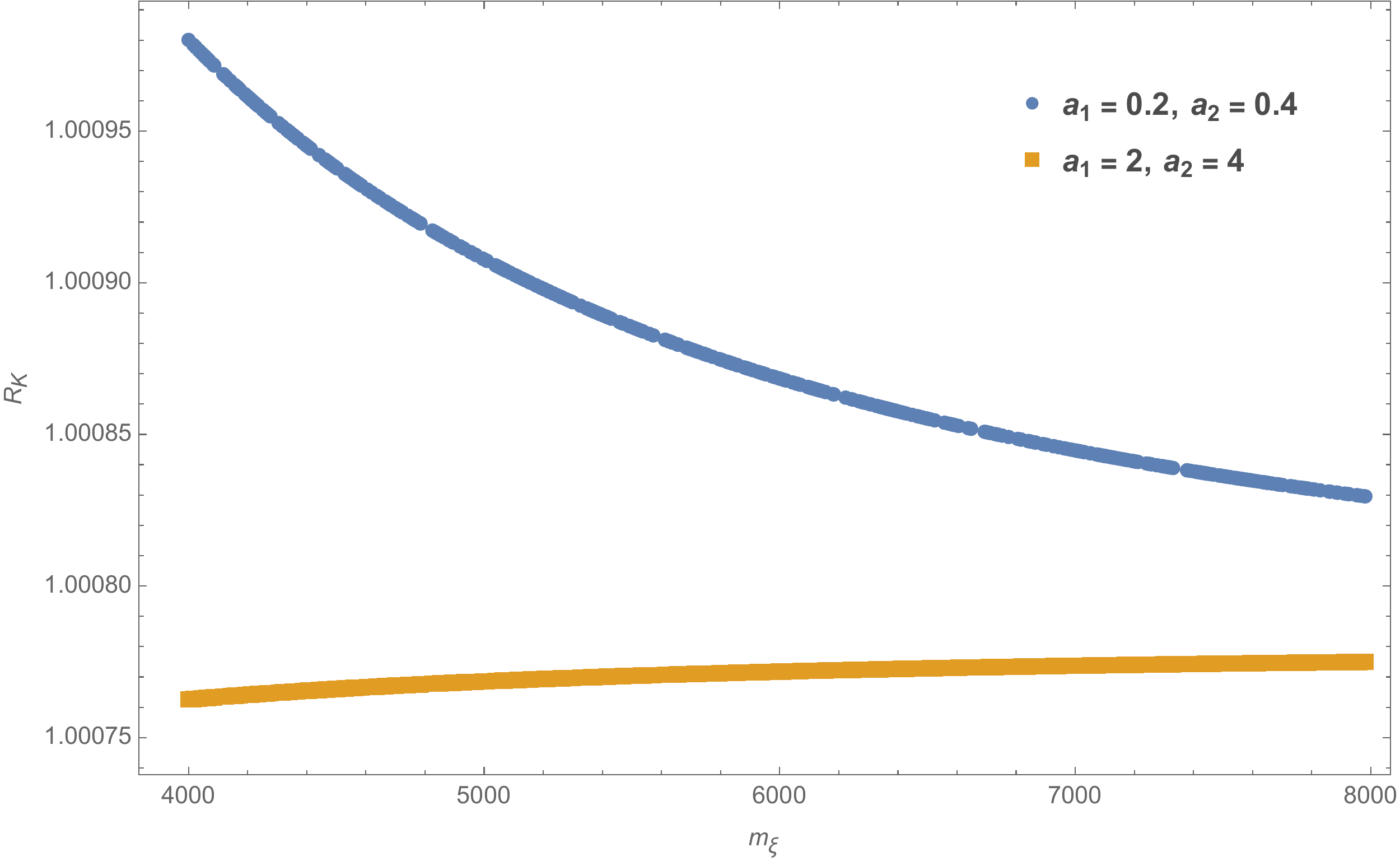}
	\caption{\label{RKbb}Ratio $\text{R}_\text{K}$ as a function of new fermion mass in the case of non-degenerate masses.} 
\end{figure}
In the second scenario, for non-degenerate masses, we plot $\text{R}_\text{K}$ as a function of new fermion 
mass. The predicted result given in Fig.(\ref{RKbb}) is similar to that of the SM, due to the suppression of factors
$\fr{m_W^2}{m_X^2}$ in Wilson coefficients $\Delta C_{9,10}^{\text{box}-\text{e}}$ and $\fr{m_Z^2}{m_{Z'}^2}$ in $\Delta C_{9,10}^{e,\mu,Z'}$. This means that in this circumstance, the ratio $\text{R}_\text{K}$ can not reach the experimental value.

To investigate the role of each source, we turned off one of the two LFUV sources. When the source from box diagrams for the first lepton generation $\Delta C_9^e$ and $\Delta C_{10}^e$ is turned off, we study $\text{R}_\text{K}$. The $\text{R}_\text{K}$
is caused by 
non-LFUV couplings of new neutral gauge boson $\text{Z}^{\prime}$ with leptons, $\text{g}^{\text{Z}^\prime}(e)\neq \text{g}^{\text{Z}^\prime}(\mu,\tau)$. As a result, the $\text{R}_\text{K}$ is determined by the mass of new $Z^\prime$ bosons, new quarks. Fig. (\ref{RKnobox}) 
plays the relationship between $\text{R}_\text{K}$ and the new quark mass $\text{m}_{\text{U}}$ via fixing the $\text{Z}'$ mass.
If the mass of the new gauge boson $\text{m}_{\text{Z}^\prime} =500$ GeV, the $\text{R}_\text{K}$ ratio can reach the experimental value, and if $\text{m}_{\text{Z}^\prime} =4000$ GeV, the ratio approaches 1. The LHC's limitations, on the other hand, suggest that the lower limit of $Z$'s mass is few TeV, which nearly closes the value of SM prediction $R_K^{\text{SM}}\simeq 1$. This means that the contribution of box diagrams for first lepton generation is a relevant source to explain  $\text{R}_\text{K}$ anomaly. So we conclude that the  $\text{R}_\text{K}$  puzzle in the MF331 model can only be solved if there is both degeneration in masses of new particles and box diagram contributions for only first generation leptons. 

\begin{figure}[H]
	\centering
	\includegraphics[width=0.7\textwidth]{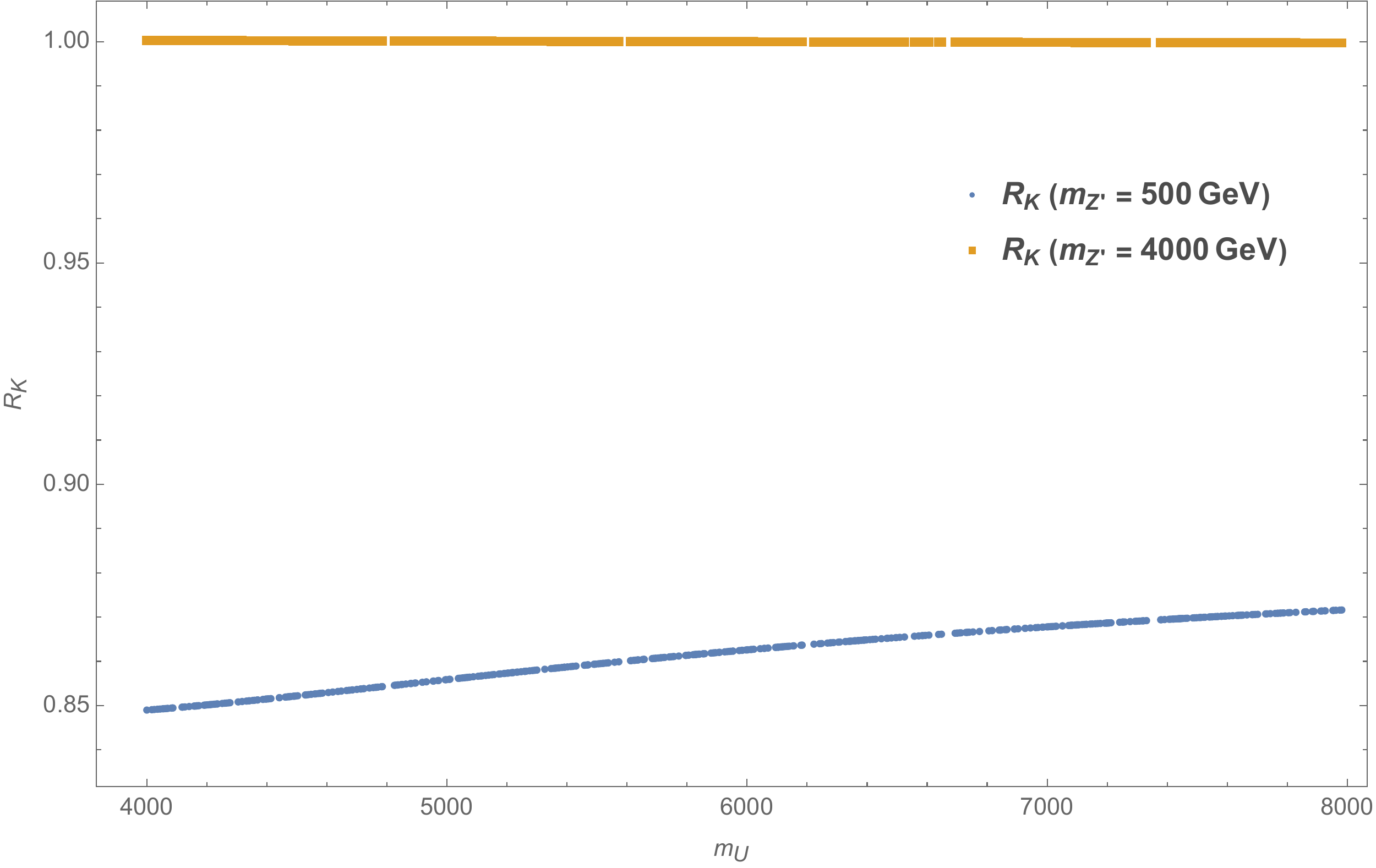}
	\caption{\label{RKnobox}The ratio $\text{R}_\text{K}$ as a function of exotic quark mass when the box diagram contributions  $\Delta \text{C}_{9,10}^{\text{box}-\text{e}}$ are excluded.}
\end{figure}

\subsection{Lepton non-universality in $\text{B}^0 \rightarrow \text{K}^{0*} \text{l}^+ \text{l}^-$}
Studying the $\text{B}^{0}\rightarrow \text{K}^{0*}\text{l}^+\text{l}^-$ decay is more complicated than studying the last process, $\text{B}^{+}\rightarrow \text{K}^{+}\text{l}^+\text{l}^-$, due to the polarization of the daughter particle $\text{K}^{0*}$ meson. The differential decay rate of $\text{B}^0\rightarrow \text{K}^{0*} \text{l}^{+}\text{l}^{-}$ can be expressed as the sum of longitudinal and transverse polarization components \cite{MunirBhutta:2020ber} using the notation employed in Eq. (\ref{Br1}) and keeping the lepton mass $\text{m}_\text{l}$
\bea
&& \fr{d\Ga(\text{B}^0\rightarrow \text{K}^{0*}\text{l}^+\text{l}^-)}{\text{dq}^2}=\fr{d\Ga_{\text{L}}(\text{B}^0\rightarrow \text{K}^{0*}\text{l}^+\text{l}^-)}{\text{dq}^2}+\fr{d\Ga_{\text{T}}(\text{B}^0\rightarrow \text{K}^{0*}\text{l}^+\text{l}^-)}{\text{dq}^2}, \crn
&& \fr{d\Ga_{\text{L}}(\text{B}^0\rightarrow \text{K}^{0*}\text{l}^+\text{l}^-)}{\text{dq}^2}= \fr{\Ga_0 \text{q}^2\sqrt{\la^{*}}\beta_\text{l}}{3}\left[\fr{3(1-\beta_\text{l}^2)}{2}(|\text{H}_\text{t}^2|^2+|\text{H}_{0}^1|^2)+\beta_\text{l}^2(|\text{H}_{0}^1|^2+ |\text{H}_{0}^2|^2)\right], \crn  
&&\fr{d\Ga_{\text{T}}(\text{B}^0\rightarrow \text{K}^{0*}\text{l}^+\text{l}^-)}{\text{dq}^2}=\fr{\Ga_0 \text{q}^2\sqrt{\la^{*}}\beta_\text{l}}{3}\sum_{\text{i}=\pm}\left[\fr{3(1-\beta_\text{l}^2)}{2}|\text{H}_\text{i}^1|^2+\beta_\text{l}^2(|\text{H}_\text{i}^1|^2+ |\text{H}_\text{i}^2|^2)\right] \label{BrBtoKs},
\eea
where helicity amplitudes $\text{H}_\text{i}^{1,2}, i=0,\pm,t$ have the following forms 
\bea
&& \text{H}_\text{t}^2=-i[\text{C}_{10}^{\text{SM}}+\Delta \text{C}_{10}]\sqrt{\fr{\la^*}{\text{q}^2}}\text{A}_0(\text{q}^2),\crn 
&& \text{H}_{\pm}^1=[\text{C}_9^{\text{eff-SM}}(\text{q}^2)+\Delta \text{C}_9]\left[\pm i\sqrt{\la^*}\fr{\text{V}(\text{q}^2)}{\text{m}_{\text{B}^0}+\text{m}_{\text{K}^{0*}}}-i(\text{m}_{\text{B}^0}+\text{m}_{\text{K}^{0*}})\text{A}_1(\text{q}^2)\right]
\crn && +\fr{2\text{m}_\text{b}}{\text{q}^2}[\text{C}_7^{\text{eff-SM}}(\text{q}^2)+\Delta \text{C}_7]\left[\pm i\sqrt{\la^*}\text{T}_1(\text{q}^2)-i(\text{m}_{\text{B}^0}^2-\text{m}_{\text{K}^{0*}}^2)\text{T}_2(\text{q}^2)\right], \crn 
&& \text{H}_{\pm}^2=(\text{C}_{10}^{\text{SM}}+\Delta \text{C}_{10})\left[\pm i\sqrt{\la^*}\fr{\text{V}(\text{q}^2)}{\text{m}_{\text{B}^0}+\text{m}_{\text{K}^{0*}}}-i(\text{m}_{\text{B}^0}+\text{m}_{\text{K}^{0*}})\text{A}_1(\text{q}^2)\right], \crn 
&& \text{H}_{0}^1=-i\fr{8\text{m}_{\text{B}^0}\text{m}_{\text{K}^{0*}}}{\sqrt{q^2}}\left[(\text{C}_9^{\text{eff-SM}}(\text{q}^2)+\Delta \text{C}_9)\text{A}_{12}(\text{q}^2)+\fr{\text{m}_\text{b}}{\text{m}_{\text{B}^0}+\text{m}_{\text{K}^{0*}}}(\text{C}_7^{\text{eff-SM}}(\text{q}^2)+\Delta \text{C}_7)\text{T}_{23}(\text{q}^2)\right], \crn 
&& \text{H}_{0}^2=-i\fr{8\text{m}_{\text{B}^0}\text{m}_{\text{K}^{0*}}}{\sqrt{\text{q}^2}}[\text{C}_{10}^{\text{SM}}+\Delta \text{C}_{10}]\text{A}_{12}(\text{q}^2),  
\eea
with $\la^*$ is the function of $\text{m}_{\text{B}^0},\text{m}_{\text{K}^{0*}}$, and $\text{q}^2$
\bea  \la^{*}(\text{q}^2)=\text{m}_{\text{B}^0}^4+\text{m}_{\text{K}^{0*}}^4+\text{q}^4-2(\text{m}_{\text{B}^0}^2\text{m}_{\text{K}^{0*}}^2+\text{q}^2\text{m}_{\text{B}^0}^2+\text{q}^2\text{m}_{\text{K}^{0*}}^2).\eea 
The non-factorizable contributions in the region $\text{q}^2\in [0,6] \ \text{GeV}^2$ can be included in the calculation by modifying the $\text{C}^{\text{eff-SM}}_7(\text{q}^2)$ \cite {Bharucha:2015bzk} as follow
\bea \text{C}_7^{\text{eff-SM}}(\text{q}^2)\rightarrow \text{C}_7^{\text{eff-SM}}(\text{q}^2)\left[1+a_i e^{i\phi_{\text{a}_i}}+\text{b}_i e^{i\phi_{\text{b}_i}}\left(\fr{\text{q}^2}{6 \ \text{GeV}^2}\right)\right], \eea 
where the ranges of the parameters $\text{a}_i,\text{b}_i$ are as follows : $\text{a}_{\pm} \in [0,0.05],\text{b}_{\pm}\in [0,0.2],\text{a}_\text{0} \in [0,0.2],\text{b}_\text{0}\in [0,0.05]$. The strong phases $\ph_{\text{a}_\text{i},\text{b}_\text{i}}$ are the  same as in the $\text{B}^+ \rightarrow \text{K}^{+}\text{l}^+\text{l}^-$ decay. The seven form factors $\text{F}_\text{i}=\text{V},\text{T}_{\text{1,2,23}},\text{A}_{\text{0,12}}$ are given in  \cite{Bharucha:2015bzk} 
\bea 
\text{F}_\text{i}(\text{q}^2)=\fr{1}{1-\text{q}^2/\text{m}_{\text{R,i}}^2}\sum_{\text{k=0,1,2}}\al_\text{k}^\text{i} [\text{z}'(\text{q}^2)-\text{z}'(0)]^\text{k}, \hs  \text{z}'(\text{q}^2)=\fr{\sqrt{\text{t}'_{+}-\text{q}^2}-\sqrt{\text{t}'_{+}-\text{t}'_0}}{\sqrt{\text{t}'_{+}-\text{q}^2}+\sqrt{\text{t}'_{+}-\text{t}'_0}}, \eea 
with $\text{t}'_{+}=(\text{m}_{\text{B}^0}+\text{m}_{\text{K}^{\text{0*}}})^2$,$\text{t}'_0=(\text{m}_{\text{B}^0}+\text{m}_{\text{K}^{0*}})(\sqrt{\text{m}_{\text{B}^0}}-\sqrt{\text{m}_{\text{K}^{0*}}})^2$. The masses of resonances $\text{m}_{\text{R,i}}$ and coefficients $\al_\text{k}^\text{i}$ can be found respectively in Table 3 and Table 15 of \cite{Bharucha:2015bzk}. The ratio $\text{R}_{\text{K}^*}$ in the dilepton invariant mass-squared range $\text{q}^2 \in [\text{q}^2_{\text{min}},\text{q}^2_{\text{max}} ]$ is determined by 
\bea \text{R}_{\text{K}^\text{*}}=\fr{\int^{\text{q}^2_{\text{max}}}_{\text{q}^2_{\text{min}}}\fr{\text{d}\Ga(\text{B}^{0}\rightarrow \text{K}^{0*}\mu^{+}\mu^{-})}{\text{dq}^2}\text{dq}^2}{\int^{\text{q}^2_{\text{max}}}_{\text{q}^2_{\text{min}}}\fr{\text{d}\Ga(\text{B}^{0}\rightarrow \text{K}^{0*}\text{e}^{+}\text{e}^{-})}{\text{dq}^2}\text{dq}^2}.
\eea
As mentioned previously, LHCb has confirmed the ratio $\text{R}_{\text{K}^{*}}$ in the dilepton invariant mass-squared range $\text{q}^2\in[1.1,6]\ \text{GeV}^2$ : $\text{R}_{\text{K}^{*}}^{\text{LHCb}}=0.685^{+0.113}_{-0.069}\pm0.047$ \cite{Aaij_2017} which yields approximately 2.5 $\sigma$ deviation from the SM predictions. We will now investigate numerically for the ratio $\text{R}_{\text{K}^*}$ predicted by the \text{MF331} model using input parameters listed in Table \ref{input1}

\begin{figure}[H]
 \begin{center}
		\includegraphics[width=0.7\textwidth]{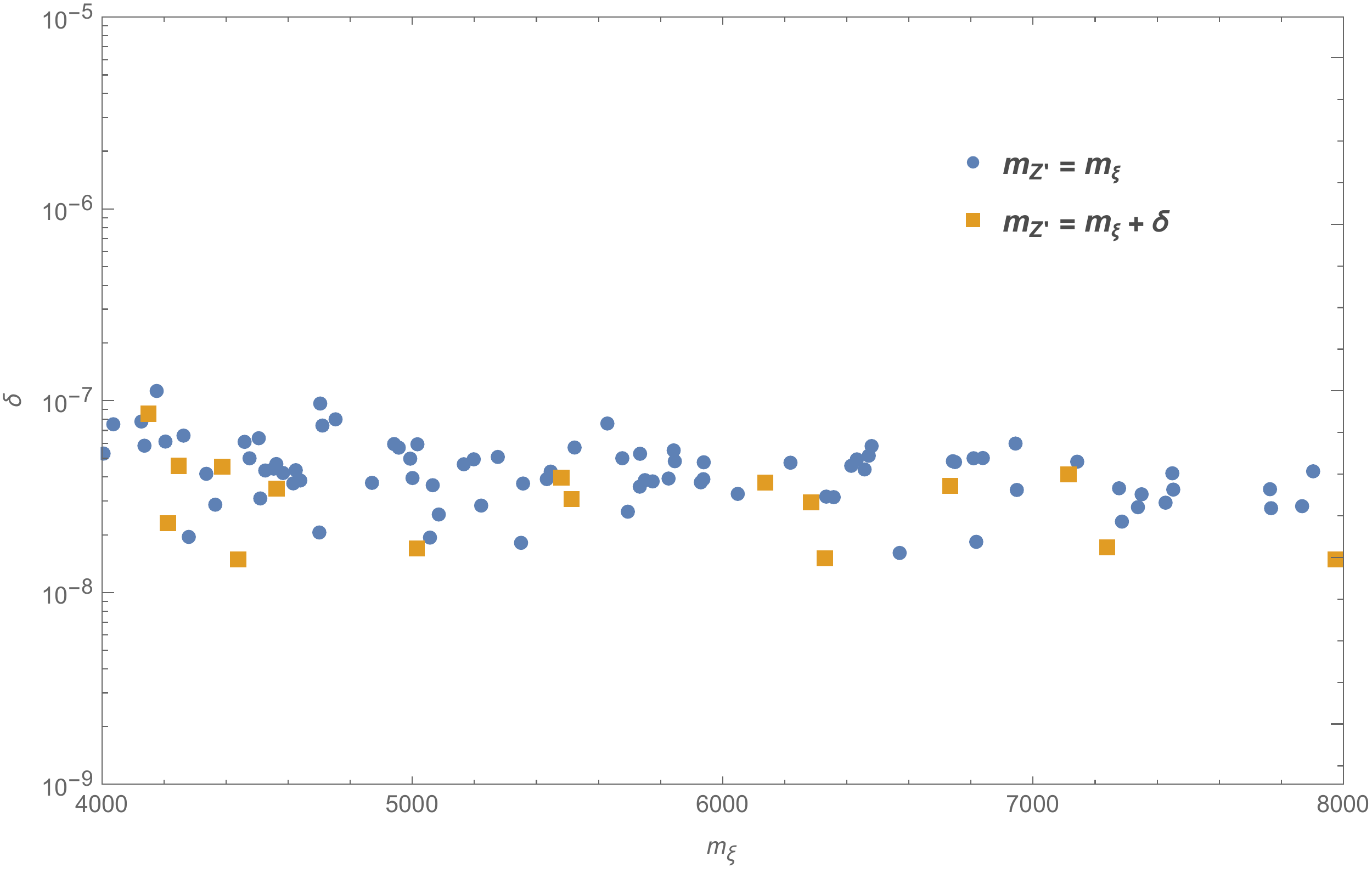}
		\end{center}
	\caption{\label{RKs}The viable regimes (blue, orange) obtained from 
		the most recent measurement \cite{Aaij_2017}, $\text{R}_{\text{K}^{*}}^{\text{LHCb}}\left(\left[1.1, 6 \right] \text{GeV}^2 \right)= 0.685^{+0.113}_{-0.069}\pm0.047$. Here $m_U=m_{\zeta^0}+\delta$ }
\end{figure}
Fig. (\ref{RKs}) shows the parameter region satisfying the experimental constraints, $\text{R}_{\text{K}^{*}}^{\text{LHCb}}\left(\left[1.1, 6 \right] \text{GeV}^2 \right)= 0.685^{+0.113}_{-0.069}\pm0.047$ by seeding parameters like $\text{m}_{\zeta^0}, \delta$ at random in the range $\text{m}_{\zeta^0} \in [4000,8000]$ GeV, $\delta \in [10^{-8},10^{-5}]$. The obtained parameters shown in the Fig. (\ref{RKs}) overlaps with the parameter domain obtained by the constraint of the $\text{R}_{\text{K}}$ measurement. Fig. (\ref{RK-RKs}) shows the image for a more accurate assessment.

\begin{figure}[H]
	\centering
	\begin{minipage}{0.5\textwidth}
	\includegraphics[width=1\textwidth]{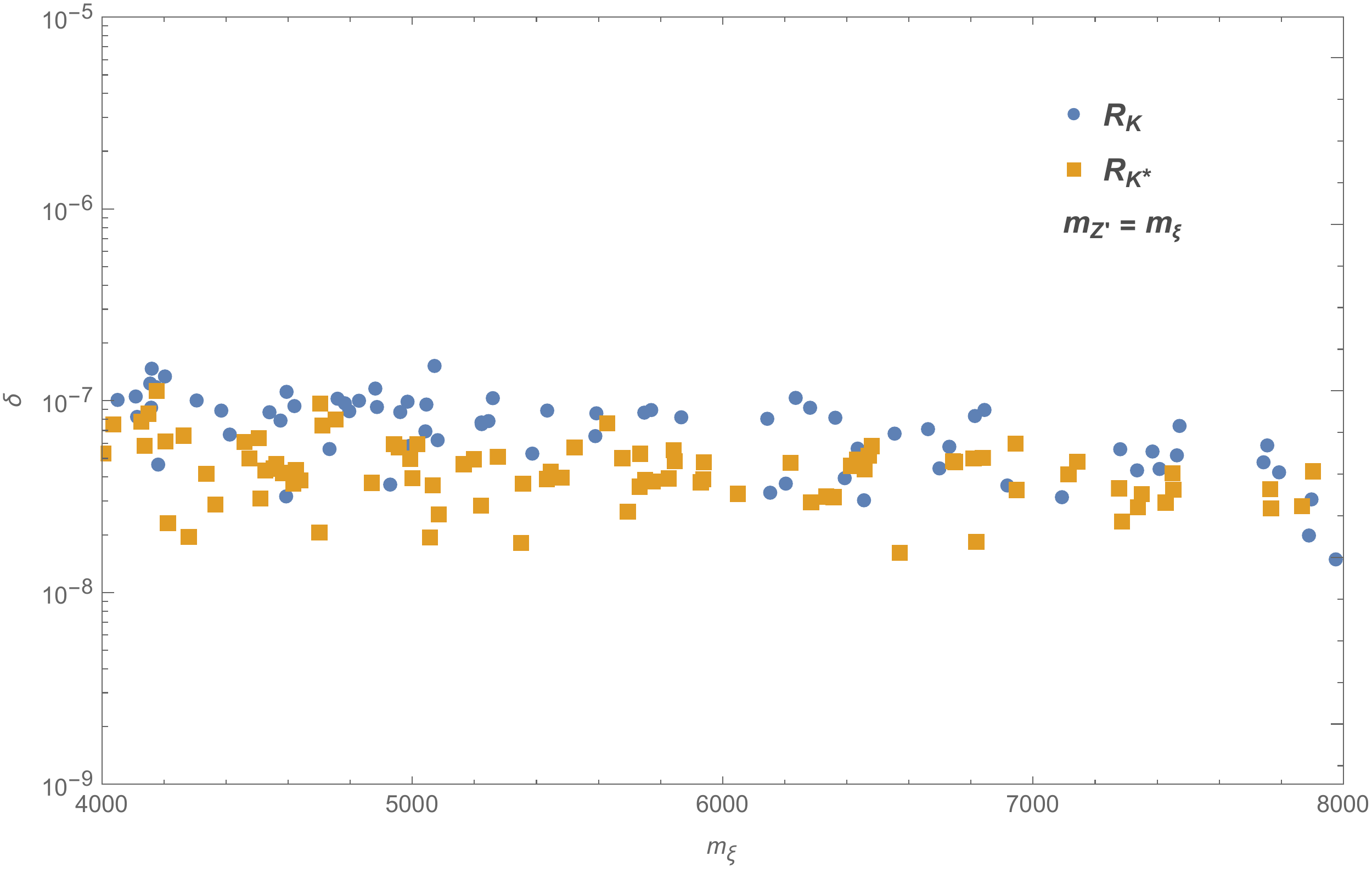}
	\end{minipage}\hfill
	\begin{minipage}{0.5\textwidth}
	\includegraphics[width=1\textwidth]{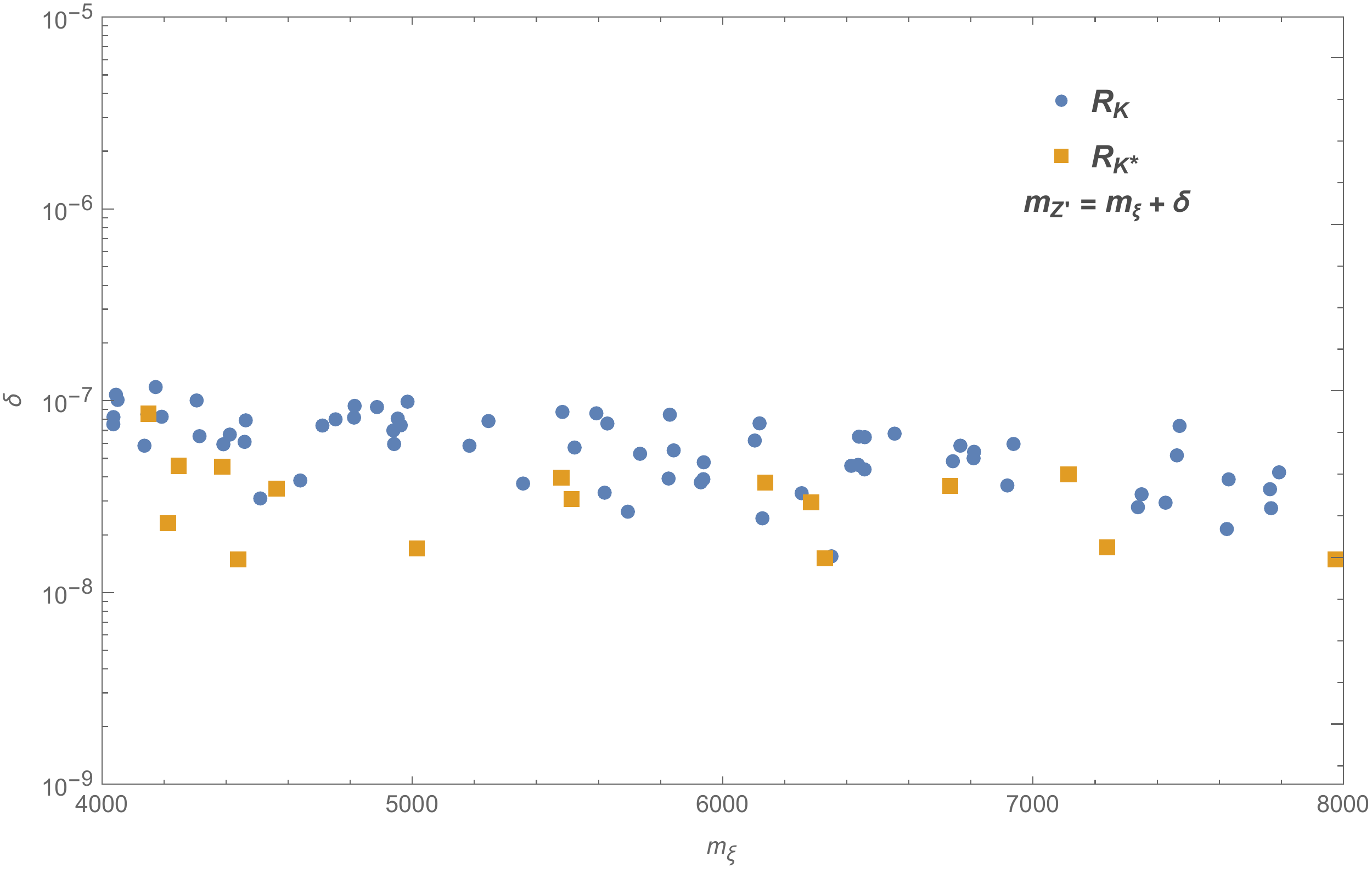}
	\end{minipage}\par
	\caption{\label{RK-RKs}The left and right panel show regions for both experimental values $\text{R}_\text{K}^{\text{LHCb}}([1.1,6] \ \text{GeV}^2)$ \cite{Aaij2021} and $\text{R}^{\text{LHCb}}_{\text{K}^*}([1.1,6] \ \text{GeV}^2)$ \cite{Aaij_2017}, with $\text{m}_{\text{Z}'}=m_{\zeta^0}$ and $\text{m}_{\text{Z}'}=m_{\zeta^0}+\delta$, respectively.}
\end{figure}

\section{\label{Bsmumu}The decay $\text{B}_s \rightarrow \mu^+ \mu^-$}
Among the non-LFUV observables, the \text{Br}$\left(\text{B}_s \rightarrow \mu^+ \mu^- \right)$ is one of the clean observables and sensitive to physics beyond the SM. This kind of decay gives a good handle on the muon sector without involving the electron sector. The theoretical prediction for branching ratio, $\text{Br}^{\text{th}}\left( \text{B}_\text{s} \rightarrow \mu^+ \mu^-\right)$, of this process remains as follows:
\bea
\text{Br}\left( \text{B}_{s} \rightarrow \mu^+ \mu^-\right)^{\text{th}}=
\tau_{\text{B}_s}\frac{\al^2 \text{G}_\text{F}^2 \text{m}_{\text{B}_s}}{16 \pi^3}\times \sqrt{1-\frac{4\text{m}_\text{l}^2}{\text{m}^2_{\text{B}_s}}}|\text{V}_{\text{tb}}\text{V}_{\text{ts}}^*|^2\text{f}^2_{\text{B}_s}\text{m}_\text{l}^2|\text{C}_{10}^{\text{SM}}+\Delta \text{C}_{10}|^2.\eea




If including the effects of $\text{B}_s-\bar{\text{B}}_s$ oscillations, the $\text{Br}\left( \text{B}_{s} \rightarrow \mu^+ \mu^-\right)^{\text{th}}$ relates  to the available experimental value as 
\bea
\text{Br}\left(\text{B}_{s} \rightarrow \mu^+ \mu^-\right)^{\text{exp}} \simeq \frac{1}{1-\text{y}_s}\text{Br}\left( \text{B}_{s} \rightarrow \mu^+ \mu^-\right)^{\text{th}},
\eea
where \bea \text{y}_s =\frac{\Delta \Gamma_{\text{B}_s}}{2\Gamma_{\text{B}_s}}=0.0645(3), \hs 
\text{Br}(\text{B}_s\rightarrow \mu^+ \mu^-)^{\text{exp}}=(3.09^{+0.46\ +0.15}_{-0.43 \ -0.11 } ) \times 10^{-9}, \label{Bmumua}\eea
experimentally established by the LHCb Collaboration \cite{Amhis:2019ckw}.

This experimental upper bound is close to the SM expectation for the  $\text{Br}\left( \text{B}_\text{s} \rightarrow \mu^+ \mu^-\right)$ \cite{Bobeth:2013uxa} (including the effect of $\text{B}_s-\bar{\text{B}}_s$ oscillations)  
\bea \text{Br}\left(\text{B}_s \rightarrow \mu^+ \mu^- \right)^{\text{SM}}=\left(3.65 \pm 0.23 \right) \times 10^{-9}.\eea 
These results impose considerable limits on the NP scale. We use numerical analysis further to investigate this constraint on the parameters in MF331.  
The NP regimes that satisfy the limits (\ref{Bmumua}) are shown by the green regions in Fig. (\ref{Bmumu}). The blue lines are contours for the experiment's center value. At least one of the exotic quarks and new gauge bosons must have a mass of a few TeV to reach the experiment's center value. Despite the fact that the NP scale is only a few TeV, NP's contribution does not exceed the boundaries (\ref{Bmumua}). 

\begin{figure}[H]
	\centering	
	\includegraphics[width=0.65\textwidth]{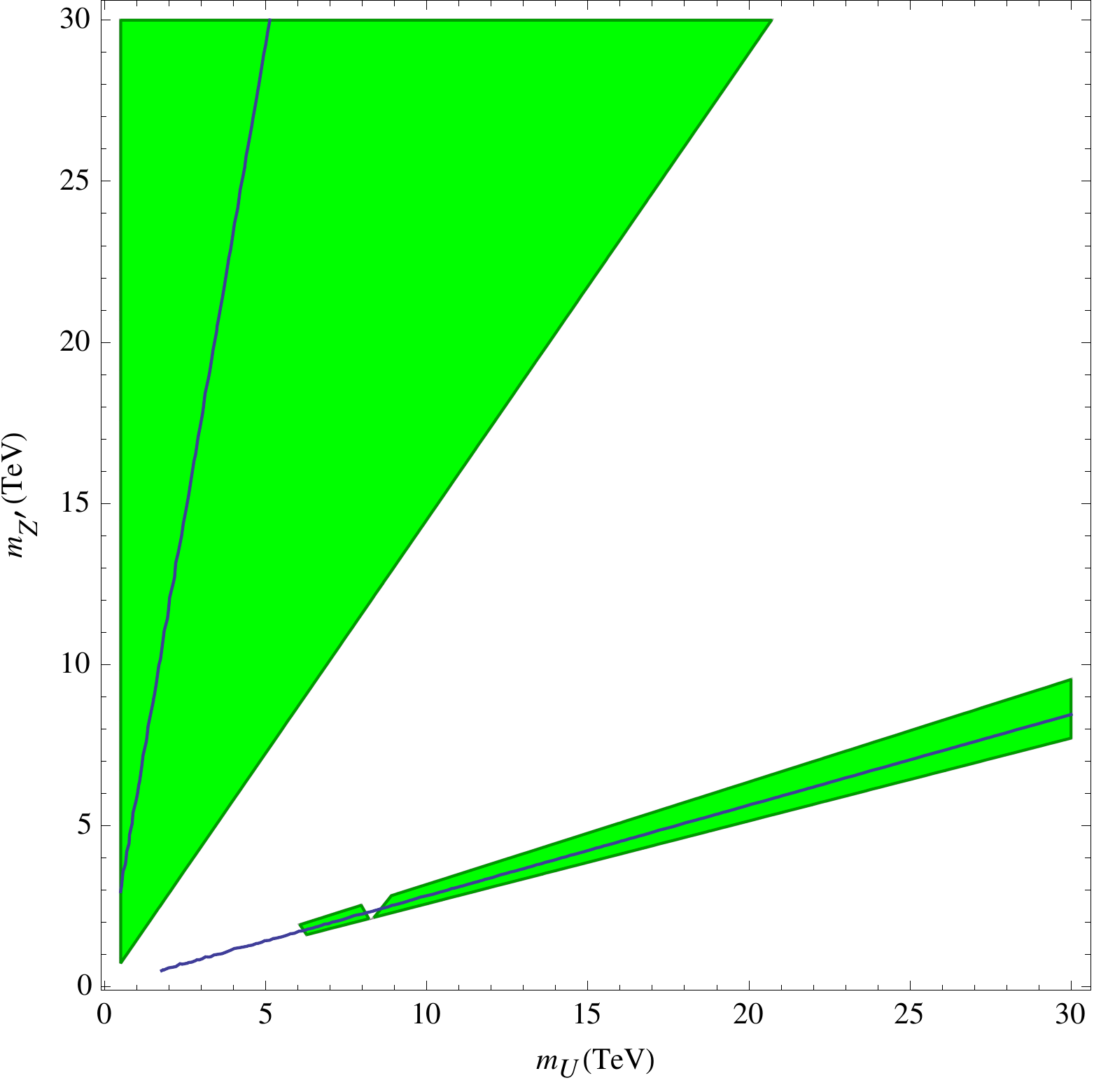}	
	\caption{\label{Bmumu} The viable NP regimes (green) as determined by the most recent measurement \cite{Amhis:2019ckw}, $\text{Br}(\text{B}_s\rightarrow \mu^+ \mu^-)^{\text{exp}}=(3.09^{+0.46\ +0.15}_{-0.43 \ -0.11 } ) \times 10^{-9}$. The blue lines present the contour for the central value of the measurement.}
	\end{figure}
\section{\label{bsgamma} The decay $\text{b} \rightarrow \text{s} \gamma$ }
The  $\text{b} \rightarrow \text{s}\gamma$ decay is also of great interest. The calculation of branching ratio at $\text{NNLO}$ level in the SM is shown in \cite{PhysRevLett.114.221801},
$\text{Br}\left(\text{b} \rightarrow \text{s}\gamma \right)^{\text{SM}}= \left(3.36 \pm 0.23\right) \times 10^{-4}$ for the photon energy $\text{E}_\ga > 1.6  \text{GeV}$ in the decay meson rest. This result is quite consistent with the combined result from measurements of CLEO, Belle, and BABAR gains \cite{Amhis:2019ckw}, $\text{Br}\left(\text{b} \rightarrow \text{s}\gamma \right)^{\text{exp}}= \left(3.32 \pm 0.15\right) \times 10^{-4}$. These findings have the potential to place strong limitations on the NP scale. New contributions to the $\text{b} \rightarrow \text{s} \ga$ decay are induced at the one-loop level in most of
NP scenarios by charged currents connected to new charged particles (new gauge boson, charged Higgs), and FCNCs associated to new neutral gauge bosons \cite{Blanke_2012,duy:2021physical}. As stated in Sec.(\ref{MF331}), the MF331 does not exist charged Higgs and FCNCs, the new effects in the $\text{b} \rightarrow \text{s} \ga$ decay are induced by the charged currents connected to gauge boson $\text{X}^\pm$. Compared with the contribution of the SM, the new contribution is strongly suppressed by a factor  $\frac{\text{m}_\text{W}^2}{\text{m}_{\text{X}}^2}$. Because scalar charged currents are not present, the new effects in $\text{b} \rightarrow \text{s} \ga$ decay in the MF331 model are expected to be minor. 

\begin{figure}[H]
	\centering
	\includegraphics[width=0.75\textwidth]{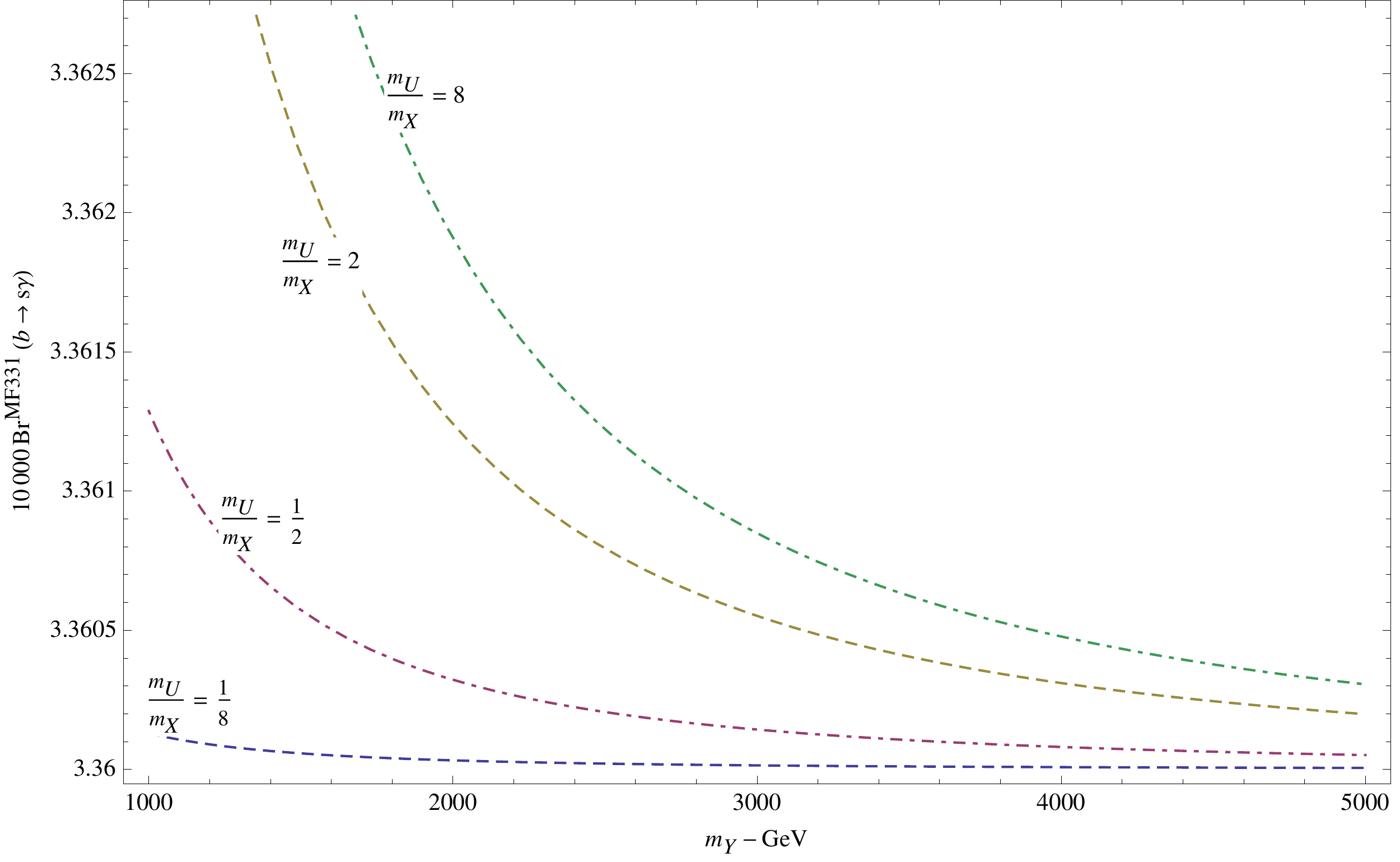}
	\caption{\label{bsga} The $\text{Br}\left(\text{b} \rightarrow \text{s}\gamma \right)$ as a function of the $\text{m}_{\text{X}}$ for fixing $\frac{\text{m}_\text{U}}{\text{m}_\text{X}}$}.
\end{figure}
Fig.(\ref{bsga}) plays the $\text{Br}\left(\text{b} \rightarrow \text{s}\gamma \right)^{\text{MF331}}$ including the NLO QCD corrections. The NP has a minor influence on the $\text{Br}\left(\text{b} \rightarrow \text{s}\gamma \right)^{\text{MF331}}$,
because the contributions of NP to the $\text{Br}\left(\text{b} \rightarrow \text{s}\gamma \right)^{\text{MF331}}$ are suppressed by factor $\frac{\text{m}_\text{U}}{\text{m}_\text{X}}$ compared to these of the SM. The $\text{Br}\left(\text{b} \rightarrow \text{s}\gamma \right)^{\text{MF331}}$ slight enhancements to the SM's value. It means that the new contribution still ensures the $\text{Br}\left(\text{b} \rightarrow \text{s}\gamma \right)^{\text{MF331}}$  to be in agreement with the present limit of the experiment \cite{Amhis:2019ckw}.

\section{\label{concl} Conclusions}
The MF331 model naturally breaks the LFU at the tree-level because the first lepton family transforms differently than the remaining lepton families. The FCNC in the quark sector does not exist at the tree-level but is allowed at the loop level due to three quark families transforming identically under the gauge symmetries. Thus, the coupling of new neutral gauge boson $Z^\prime$ with a pair of $\text{e}^+ \text{e}^-$ differs from that of $\mu^+\mu^-$ and $\tau^+ \tau^-$, whereas three quark families couples with the same strength to the $\text{Z}^\prime$-boson. Based on this feature, we investigate contributions from the $\gamma,\text{ Z}, \text{Z'}-$penguin diagrams to the b-s transitions and, combining them with the tree-level interactions of $\ga, \text{Z}, \text{Z}^\prime$ with a pair of leptons, we induce the NP contributions to the WCs $\Delta C_{7,8,9,10}^{e,\mu,\tau}$. The $\gamma,\text{ Z}-$penguin diagrams give the 
same contributions to the WCs for three generations of leptons, but the $\text{Z}^\prime-$penguin diagrams give different contributions between the first lepton generation and the other two generations. Another interesting feature of the model is that  first family's left-handed leptons are classified as hexagons of the $SU(3)_L$ group, resulting in the appearance of new leptons, $\xi^0, \xi^\pm$. The newly charged lepton current, $\bar{\xi}^0\ga^\mu e$, couples to the newly charged gauge boson $X^+_\mu$, resulting in a box diagram that only shows
the first lepton family's contribution to the WCs.  That's why the MF331 model provides two possible sources of contributions to non-LFU effective interactions, which allow us to explain the $\text{R}_{\text{K}}, \text{R}_{\text{K}^*}$ anomalies. The  $\text{Z'}-$penguin diagrams give a negligible contribution by comparing to the SM contributions, because the contribution of $\text{Z}^{\prime}$ is suppressed by a factor $ \fr{\text{m}_\text{Z}^2}{\text{m}_{\text{Z}^\prime}^2}$. We show that the $\text{R}_{\text{K}}, \text{R}_{\text{K}^*}$ anomalies can be explained by contributing from the box diagrams in the situation of a mass degeneracy of the new particle. In the allowed region of the NP scale, we investigate the NP contributions to the $\text{Br}(B_s \to \mu^+ \mu^-), \text{Br}(b \to s \gamma)$. These contributions are consistent with the experimental measurements.
\section*{Acknowledgments}\vspace{-0.4cm}
This research is funded by Vietnam Academy of Science and Technology under Grant No. NVCC05.09/22-23.
\centerline{\bf{REFERENCES}}\vspace{-0.4cm} 
\bibliographystyle{utphys}
\bibliography{Refflip11}

\end{document}